\documentclass[twocolumn,showpacs,preprintnumbers,superscriptaddress,amsmath,amssymb,prd,floatfix]{revtex4}%
\usepackage{graphicx} % Include figure files
\usepackage{dcolumn}  % Align table columns on decimal point
\usepackage{amsmath,amssymb}
\usepackage{color,rotating}
\usepackage{epic,eepic}
\usepackage{fancybox}
\usepackage{psboxit}
\usepackage{axodraw}
\usepackage{pstricks,pst-node,pst-text}%,pst-osci}
\usepackage{charter}

\graphicspath{{ps}}

\begin{document}

\preprint{\vbox{ \hbox{   }
    \hbox{Belle Preprint 2008-14}
    \hbox{KEK Preprint 2008-8} 
%    \hbox{B. Golob(chair),W. Bartel, Kaji} 
%    \hbox{hep-ex nnnn, if available}
}}

\title {Measurement of Azimuthal Asymmetries in Inclusive Production\\of Hadron Pairs in $e^+e^-$ Annihilation at $\sqrt{s}=10.58$ GeV}

\date{\today}% It is always \today, today,
\begin{abstract}
%\linenumbers
The Collins effect connects transverse quark spin with a
measurable azimuthal asymmetry in the yield of hadronic fragments
around the quark's momentum vector. Using two different reconstruction
methods we measure statistically significant azimuthal asymmetries for
charged pion pairs in $e^+e^-$ annihilation at center-of-mass energies
of 10.52 GeV and 10.58 GeV, which can be attributed to the fragmentation of primordial quarks with transverse spin components. The measurement was performed using a dataset of 547 fb$^{-1}$ collected by the Belle detector at KEKB improving the statistics of the previously published results by nearly a factor of 20. 
\end{abstract}

%%% Paper:    Azimuthal asymmetries in e+e- -> h h
%%% Journal:  Physical Review D
%%% Contacts: R. Seidl (rseidl@uiuc.edu)
%%%           M. Grosse-Perdekamp (mgp@uiuc.edu)
%%%           A. Ogawa (akio@rcf2.rhic.bnl.gov)
%%% Non-responding authors or those who said NO are commented out.
%%% ====================================================================
%%% Click the RELOAD button on your web browser to see the updated file.
%%% ====================================================================
%%% Use \input{author} to insert this material into your latex file.
%%%%% Force institutions to appear in alphabetical order when typeset.
\affiliation{Budker Institute of Nuclear Physics, Novosibirsk}
\affiliation{Chiba University, Chiba}
\affiliation{University of Cincinnati, Cincinnati, Ohio 45221}
\affiliation{T. Ko\'{s}ciuszko Cracow University of Technology, Krakow}
\affiliation{Deutsches Elektronen--Synchrotron, Hamburg}
%%%\affiliation{Department of Physics, Fu Jen Catholic University, Taipei}
\affiliation{Justus-Liebig-Universit\"at Gie\ss{}en, Gie\ss{}en}
\affiliation{The Graduate University for Advanced Studies, Hayama}
%%%\affiliation{Gyeongsang National University, Chinju}
\affiliation{Hanyang University, Seoul}
\affiliation{University of Hawaii, Honolulu, Hawaii 96822}
\affiliation{High Energy Accelerator Research Organization (KEK), Tsukuba}
%%%\affiliation{Hiroshima Institute of Technology, Hiroshima}
\affiliation{University of Illinois at Urbana-Champaign, Urbana, Illinois 61801}
\affiliation{Institute of High Energy Physics, Chinese Academy of Sciences, Beijing}
\affiliation{Institute of High Energy Physics, Vienna}
\affiliation{Institute of High Energy Physics, Protvino}
\affiliation{Institute for Theoretical and Experimental Physics, Moscow}
\affiliation{J. Stefan Institute, Ljubljana}
\affiliation{Kanagawa University, Yokohama}
\affiliation{Korea University, Seoul}
%%%\affiliation{Kyoto University, Kyoto}
\affiliation{Kyungpook National University, Taegu}
\affiliation{\'Ecole Polytechnique F\'ed\'erale de Lausanne (EPFL), Lausanne}
\affiliation{Faculty of Mathematics and Physics, University of Ljubljana, Ljubljana}
\affiliation{University of Maribor, Maribor}
\affiliation{University of Melbourne, School of Physics, Victoria 3010}
\affiliation{Nagoya University, Nagoya}
\affiliation{Nara Women's University, Nara}
\affiliation{National Central University, Chung-li}
\affiliation{National United University, Miao Li}
\affiliation{Department of Physics, National Taiwan University, Taipei}
\affiliation{H. Niewodniczanski Institute of Nuclear Physics, Krakow}
\affiliation{Nippon Dental University, Niigata}
\affiliation{Niigata University, Niigata}
\affiliation{University of Nova Gorica, Nova Gorica}
\affiliation{Osaka City University, Osaka}
\affiliation{Osaka University, Osaka}
\affiliation{Panjab University, Chandigarh}
%%%\affiliation{Peking University, Beijing}
%%%\affiliation{Princeton University, Princeton, New Jersey 08544}
\affiliation{RIKEN BNL Research Center, Upton, New York 11973}
\affiliation{Saga University, Saga}
\affiliation{University of Science and Technology of China, Hefei}
\affiliation{Seoul National University, Seoul}
%%%\affiliation{Shinshu University, Nagano}
\affiliation{Sungkyunkwan University, Suwon}
\affiliation{University of Sydney, Sydney, New South Wales}
%%%\affiliation{Tata Institute of Fundamental Research, Mumbai}
\affiliation{Toho University, Funabashi}
\affiliation{Tohoku Gakuin University, Tagajo}
%%%\affiliation{Tohoku University, Sendai}
\affiliation{Department of Physics, University of Tokyo, Tokyo}
%%%\affiliation{Tokyo Institute of Technology, Tokyo}
\affiliation{Tokyo Metropolitan University, Tokyo}
\affiliation{Tokyo University of Agriculture and Technology, Tokyo}
%%%\affiliation{Toyama National College of Maritime Technology, Toyama}
\affiliation{Virginia Polytechnic Institute and State University, Blacksburg, Virginia 24061}
\affiliation{Yonsei University, Seoul}
  \author{R.~Seidl}\affiliation{University of Illinois at Urbana-Champaign, Urbana, Illinois 61801}\affiliation{RIKEN BNL Research Center, Upton, New York 11973} % UIUC
  \author{M.~Grosse~Perdekamp}\affiliation{University of Illinois at Urbana-Champaign, Urbana, Illinois 61801}\affiliation{RIKEN BNL Research Center, Upton, New York 11973} % UIUC
  \author{A.~Ogawa}\affiliation{RIKEN BNL Research Center, Upton, New York 11973} % RIKEN
  \author{I.~Adachi}\affiliation{High Energy Accelerator Research Organization (KEK), Tsukuba} % KEK
  \author{H.~Aihara}\affiliation{Department of Physics, University of Tokyo, Tokyo} % Tokyo
% \author{D.~Anipko}\affiliation{Budker Institute of Nuclear Physics, Novosibirsk} % BINP
% \author{K.~Arinstein}\affiliation{Budker Institute of Nuclear Physics, Novosibirsk} % BINP
% \author{T.~Aso}\affiliation{Toyama National College of Maritime Technology, Toyama} % Toyama
% \author{V.~Aulchenko}\affiliation{Budker Institute of Nuclear Physics, Novosibirsk} % BINP
% \author{T.~Aushev}\affiliation{\'Ecole Polytechnique F\'ed\'erale de Lausanne (EPFL), Lausanne}\affiliation{Institute for Theoretical and Experimental Physics, Moscow} % ITEP
% \author{T.~Aziz}\affiliation{Tata Institute of Fundamental Research, Mumbai} % Tata
  \author{S.~Bahinipati}\affiliation{University of Cincinnati, Cincinnati, Ohio 45221} % Cincinnati
  \author{A.~M.~Bakich}\affiliation{University of Sydney, Sydney, New South Wales} % Sydney
% \author{V.~Balagura}\affiliation{Institute for Theoretical and Experimental Physics, Moscow} % ITEP
% \author{Y.~Ban}\affiliation{Peking University, Beijing} % Peking
% \author{E.~Barberio}\affiliation{University of Melbourne, School of Physics, Victoria 3010} % Melbourne
% \author{M.~Barbero}\affiliation{University of Hawaii, Honolulu, Hawaii 96822} % Hawaii
  \author{W.~Bartel}\affiliation{Deutsches Elektronen--Synchrotron, Hamburg} % DESY
% \author{A.~Bay}\affiliation{\'Ecole Polytechnique F\'ed\'erale de Lausanne (EPFL), Lausanne} % Lausanne
% \author{I.~Bedny}\affiliation{Budker Institute of Nuclear Physics, Novosibirsk} % BINP
% \author{K.~Belous}\affiliation{Institute of High Energy Physics, Protvino} % Protvino
% \author{V.~Bhardwaj}\affiliation{Panjab University, Chandigarh} % Panjab
  \author{U.~Bitenc}\affiliation{J. Stefan Institute, Ljubljana} % Ljubljana
% \author{S.~Blyth}\affiliation{National United University, Miao Li} % NUU
  \author{A.~Bondar}\affiliation{Budker Institute of Nuclear Physics, Novosibirsk} % BINP
% ...
% BINP
\collaboration{The Belle Collaboration} 
\noaffiliation
\maketitle
%\tableofcontents{}
\section{Introduction}
%\linenumbers
Electron-positron annihilation at high energies
leads to the creation of quark-antiquark pairs with high
momenta. Assuming that neither quark in a given event radiates an energetic gluon, the quark and the antiquark fragment
into two jets of hadrons in opposite hemispheres. The energies and
the momentum vectors of the quarks can be inferred from the
energies and momenta of the observed final state hadrons.
The quark-hadron fragmentation process is usually parameterized
with the help of fragmentation functions $D_q^h(z)$. These describe
number densities for a quark of flavor $q$ to fragment into a hadron
$h$ carrying the fraction $z$ of the original quark momentum.

The two final state quarks can be created in transverse spin
states. Here we use azimuthal correlations between
pairs of hadrons in opposite jet hemispheres to study transverse
spin effects in the quark fragmentation process. Specifically, we
study the distribution of final state pions around the
momentum vector of the fragmenting quark. The quark momentum
direction is measured approximately by identifying it with the
thrust axis.

Spin-dependent effects in the fragmentation of quarks into hadrons
were first discussed by Collins \cite{collins2}: the Collins function
$H_1^{\perp}(z,k_{T}^2)$ is the amplitude of the modulation in the
azimuthal distribution of the final state hadrons in spin-dependent
fragmentation processes. It depends on the normalized hadron
momentum $z$ and the magnitude of the transverse hadron momentum
$k_T$ with respect to the three momentum of the quark. Initial efforts to extract Collins asymmetries in $e^+e^-$ annihilation were carried out in DELPHI \cite{efremov}. The first
observation of the azimuthal Collins asymmetries in quark
fragmentation was reported by the Belle collaboration \cite{belleprl}. This paper is an extension of the earlier Belle
measurement with nearly a factor of 20 higher statistics allowing
a more refined analysis.

The Collins function cannot be calculated reliably employing QCD
based algorithms, because the fragmentation process is a non-perturbative
process. However, once determined experimentally the Collins effect
can be used as ``quark spin polarimeter" to determine the transverse
spin states of final state quarks in semi-inclusive deep inelastic
scattering (SIDIS) and in polarized proton-proton scattering. Over the
past few years there has been increasing interest in spin phenomena and experiments have published new data on the transverse spin structure of the nucleon \cite{hermesprl,compassprl,e704,star,brahms,phenix}. However, the semi-inclusive observables in SIDIS are not a direct measurement of the transverse quark spin distributions (the so-called transversity distributions). Instead they measure a product
of quark transversity distributions and the Collins fragmentation functions.
Only an independent measurement of the Collins asymmetry in $e^+e^-$
annihilation makes it possible to extract transversity
quark distributions from single transverse spin asymmetries in SIDIS
or polarized proton-proton scattering.

Recently, the first global transversity analysis of SIDIS and Belle
data has been carried out by Anselmino and collaborators \cite{alexei}.
The resulting $u$ and $d$ quark transversity distributions have
large errors that are dominated by the experimental
uncertainties in the SIDIS and Belle data. The new results
presented here in combination with new data expected in the near
future from HERMES and COMPASS will reduce the uncertainties
in the extraction of quark transversity distributions.

The study of transverse momentum dependent (TMD) fragmentation
and distribution functions such as the Collins fragmentation function
has led to important theoretical developments in QCD and have
greatly advanced the understanding of factorization and process
dependence of distribution- and fragmentation-functions in hard scattering
processes. TMD hard quark scattering can give rise to the large
single spin asymmetries that have been observed experimentally but
seemed to be incompatible with the traditional hard scattering
QCD framework, which assumes factorization \cite{Kane:1978nd}. Specifically, factorization assumes that the exchange of
soft gluons in the initial or final state of a hard scattering process can be neglected. Furthermore it is assumed that parton distributions and fragmentation functions are process-independent, i.e. that the Collins function in SIDIS
will be the same as in $e^+e^-$. It has now become possible to correctly
include the initial and final
state interactions in QCD calculations for some hard scattering processes.
Recent calculations \cite{Brodsky:2002cx,Bomhof:2006ra} with the correct treatment of initial and final state interactions in hard processes show that the Sivers effect \cite{Sivers} arises from this inclusion. Another consequence of this new treatment of factorization is the process dependence of the Sivers distribution function. While the Sivers distributions in SIDIS and the 
Drell-Yan process are expected to have the same magnitude they are predicted
to have opposite signs.

Collins fragmentation appears to be an
important test case for future theoretical approaches that describe low energy QCD phenomena. The Collins function describes phenomenologically the evolution into hadrons starting from a quark with a given momentum and spin orientation. The challenge for the future will be to calculate this process from basic principles.

The paper is organized as follows: after a discussion of the Collins
effect the experimental procedure will be described in detail, a
detector description is followed by a description of the analysis
and the study of systematic errors. Finally, results will be
presented together with an attempt to interpret the measured
Collins asymmetries for di-hadrons in terms of the Collins
fragmentation functions.
\subsection{Transversity}
Recently there has been an increased interest in the third leading twist quark distribution function of the nucleon, the so-called transversity distribution $\delta q(x)$ \cite{ralston}. The corresponding charge is the tensor charge, which is obtained by summing the charge-squared weighted transverse spin distribution over all flavors, integrated over the quarks' momentum fraction $x$. Due to its symmetry properties this is a more valence-like object than the axial charge (which is obtained in a similar way from the quark helicity distribution) \footnote{This aspect makes it better accessible to Lattice calculations.}.
Since the transversity distribution function is chiral-odd, gluons cannot contribute at leading order. Therefore no gluon transversity distribution exists and thus the QCD evolution of transversity is different from that of the quark helicities. Transversity cannot be accessed with inclusive DIS experiments, since the corresponding forward scattering amplitude contains a helicity flip that is suppressed by the nearly vanishing quark mass relative to the scale $\sqrt{Q^2}$. The scale in DIS is defined as $Q^2 = - q^\mu q_\mu$ where $q_\mu$ is the four-momentum transfer from the lepton to the nucleon.

As a consequence one can only measure transversity if a second chiral-odd distribution or fragmentation function is involved.
One possibility to measure transversity uses Drell-Yan processes, via double spin asymmetries of two transversely polarized proton and (anti)proton beams. This yields information on the product of a quark and an antiquark transversity distribution. Unfortunately the cross section for this process is quite small and the possibility of reasonably high transverse polarizations for antiprotons is limited although some efforts are being made to conduct such experiments in the future \cite{jparc,pax}.
Another way to access transversity is by combining it with a chiral-odd quark fragmentation function. This combination is available in semi-inclusive DIS off a transversely polarized nucleon by detecting at least one of the produced hadrons. Such measurements are currently being performed at HERMES \cite{hermesprl} on a proton and at COMPASS \cite{compassprl} on a deuteron target. 
In addition, a very similar observable can be accessed in proton-proton collisions where one of the beams is transversely polarized. In both cases there are different candidates for chiral-odd fragmentation functions, which are briefly discussed in the following three paragraphs. \\
\subsubsection{Collins fragmentation function}
Most relevant to the interpretation of existing data on experiments with transversely polarized targets or beams is the Collins function first proposed in Ref.~\cite{collins2}. The Collins function relates the transverse spin of the quark to the transverse momentum of a final state hadron. These correlations result in a non-uniform azimuthal distribution of final state hadrons around the initial quark direction. 
\subsubsection{Interference fragmentation function}
The interference fragmentation function \cite{Collins:1993kq} measures the interference of two hadrons in the final state between two different angular momentum states. This interference is again visible as an azimuthal asymmetry of the plane defined by the two hadrons around the original quark momentum. According to model predictions \cite{jaffe,radici} the interference fragmentation function may be different from zero whenever the invariant mass of the hadron pair is close to the mass of resonances. Examples are the $\rho$ for pion pairs and the $\phi$ for kaon pairs.
\subsubsection{$\Lambda$ polarimetry}
Another possibility is polarimetry of $\Lambda$ baryons produced in fragmentation \cite{mauro}. Here the relevant fragmentation function describes the fragmentation of a transversely polarized quark into a transversely polarized $\Lambda$. The $\Lambda$-polarization can be inferred experimentally from the kinematics of the decay.  \\

These three spin-dependent fragmentation processes can be used to  obtain transversity quark distributions \cite{muldersbible}. However, in all three cases the experimental observables depend on transversity distributions folded with a currently unknown spin analyzing fragmentation function. Therefore transversity functions cannot be extracted from polarized semi-inclusive measurements alone, and an independent measurement of the fragmentation functions is required. In this paper we discuss a measurement of Collins asymmetries performed at the KEKB $e^+e^-$ collider with the Belle detector. 
Extraction of transversity distributions through a combined analysis of transverse spin asymmetries from semi-inclusive polarized DIS and Collins fragmentation data from $e^+e^-$ requires the universality of the Collins fragmentation function between $e^+e^-$ and deep inelastic scattering. At the present time universality has been confirmed only at leading order by Collins and Metz \cite{Collins:2004nx}.
This paper describes the measurement of the Collins effect in $e^+e^-$ collisions.
\subsection{Collins fragmentation in $e^+e^-$ and SIDIS}
\begin{figure}[t]
\begin{center}
\includegraphics[width=7cm]{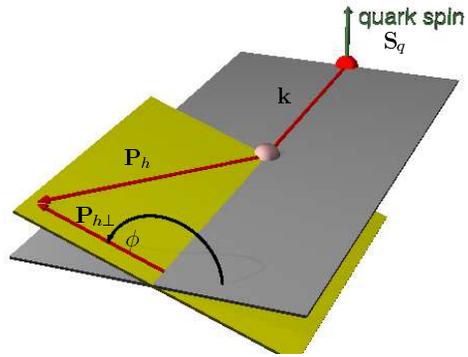}
\caption{Definition of the azimuthal angle $\phi$ in the fragmentation of a transversely polarized quark $\mathbf{S}_q$ with momentum $\mathbf{k}$ into an unpolarized hadron with transverse momentum $\mathbf{P}_{h\perp}$.}
\label{angle1}
\end{center}
\end{figure}

The Collins fragmentation function describes the creation of hadrons with transverse momentum $\mathbf{P}_{h\perp}$ from a transversely polarized quark with polarization $\mathbf{S}_q$ and momentum $\mathbf{k}$. Following the Trento convention \cite{trento} the number density for finding a hadron $h$ produced from a transversely polarized quark $q$ is defined as:
\begin{equation}
D_{h\/q^\uparrow}(z,P_{h\perp})  = D_1^q(z,P_{h\perp}^2) + H_1^{\perp q}(z,P_{h\perp}^2)\frac{(\mathbf{\hat{k}} \times \mathbf{P}_{h\perp})\cdot \mathbf{S}_q}{zM_h}\quad ,
\label{eq:cdef}
\end{equation}
where the first term is the unpolarized fragmentation function, with $z\stackrel{CMS}{=} \frac{2E_h}{Q}$ being the fractional energy of the hadron relative to the center-of-mass system (CMS) energy $\sqrt{s}=Q$. The second term contains the Collins function $H_1^{\perp q}(z,P_{h\perp}^2)$ and depends on the spin orientation of the quark. It changes sign when the quark spin is flipped and thus generates a single spin asymmetry. The vector product introduces a $\cos(\phi)$  modulation where $\phi$ is the azimuthal angle spanned by the transverse momentum and the plane normal to the quark spin along the quark's momentum (see Fig.~\ref{angle1}). 

In SIDIS one measures a product of the quark transverse polarization and the Collins function. The transverse polarization vector of the quarks serves as a
reference axis for the azimuthal asymmetries of single spin distributions. In $e^+e^-$ with unpolarized beams no such reference direction exists. An analysis of Collins asymmetries in single jets in $e^+e^-$ therefore will yield a zero result as the cosine modulation will average to zero in a large event sample. This observation will later be used for important tests of the analysis algorithm and possible false asymmetries resulting from detector effects (section \ref{sec:systematics}). The Collins effect can therefore only be observed in the combination of two functions each one creating a single spin asymmetry. The correlation of quark and antiquark Collins functions in opposite hemispheres gives a product of two $\cos(\phi)$ modulations for the two azimuthal angles $\phi_1$ and $\phi_2$, resulting in a $\cos(\phi_1+\phi_2)$  modulation (see Fig.~\ref{angle2}). In $e^+e^-$ annihilation processes these azimuthal angles are defined as
 \begin{eqnarray}
\phi_{1,2}  &= & sgn\left[ \mathbf{\hat{n}} \cdot \left\{ (\mathbf{\hat{z}}\times \mathbf{\hat{n}})  \times (\mathbf{\hat{n}} \times \mathbf{P}_{h1,2})\right\}\right] \nonumber \\
&\times & \arccos \left(\frac{\mathbf{\hat{z}}\times \mathbf{\hat{n}}}{|\mathbf{\hat{z}}\times \mathbf{\hat{n}}|}  \cdot \frac{\mathbf{\hat{n}} \times \mathbf{P}_{h1,2}}{|\mathbf{\hat{n}}\times\mathbf{P}_{h1,2}|}\right)   \quad,
\end{eqnarray}
where $\mathbf{\hat{z}}$ is a unit vector along the z-axis defined by the $e^+e^-$ beam direction and $\mathbf{\hat{n}}$ is the thrust axis, used as an approximation for the quark-antiquark axis (defined below).
\subsubsection{Transverse spin components}
\begin{figure}[t]
\begin{center}
\includegraphics[width=7cm]{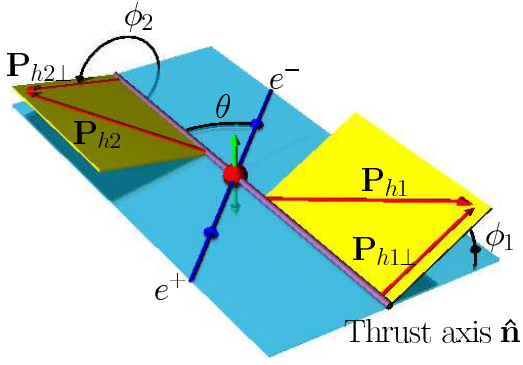}
\caption{Definition of the azimuthal angles $\phi_1$ and $\phi_2$ of the two hadrons, between the scattering plane and their transverse momenta $\mathbf{P}_{hi\perp}$ around the thrust axis $\hat{n}$. The angle $\theta$ is defined as the angle between the lepton axis and the thrust axis.}
\label{angle2}
\end{center}
\end{figure}

In $e^+e^-$ annihilation processes with unpolarized beams the spin 1 photon has equal contributions from the two lepton helicity states $|+-\rangle$ and $|-+\rangle$. In the case a quark-antiquark pair is created with a CMS angle of $\theta = \pi/2$ (see Fig.~\ref{angle2}) both lepton helicity combinations contribute equally and the transverse polarization of the quarks is zero on average. Hence the quark-antiquark pair has antiparallel transverse spin components. For finite production angles the probability to have antiparallel spins is proportional to $\sin^2\theta$. This kinematic dependence provides a powerful test of the extraction of Collins asymmetries in $e^+e^-$ and will be discussed later. 
\subsection{Azimuthal asymmetries and cross section}

Two different azimuthal asymmetries in inclusive dihadron production in $e^+e^-$ annihilation will become important in the course of the analysis. The calculation of these will be first described before discussing the cross sections. The method already mentioned in the previous subsection translates the definition of the Collins function (eq.~\ref{eq:cdef}) into the $e^+e^-\rightarrow q\bar{q}$ case, where the initial momentum of the quark-antiquark pair is known. The quark directions are, however, not accessible to a direct measurement and are thus approximated by the thrust axis. The thrust axis $\hat{n}$ maximizes the event shape variable thrust:
\begin{equation}
T \stackrel{max}{=} \frac{ \sum_h|\mathbf{P^{\mathrm{CMS}}_h}\cdot\mathbf{\hat{n}}|}{ \sum_h|P^{\mathrm{CMS}}_h|}\quad,
\end{equation}
where the sum extends over all detected particles. The thrust value varies between 0.5 for spherical events and 1 for tracks aligned with the thrust axis of an event. The thrust axis is a good approximation to the original quark-antiquark axis as described in Section \ref{sec:eventsel}. 
The first method of accessing the Collins asymmetry, $M_{12}$ is based on measuring a $\cos(\phi_1+\phi_2) $ modulation of hadron pairs ($N(\phi_1+\phi_2)$) on top of the flat distribution due to the unpolarized part of the fragmentation function. The unpolarized part is given by the average bin content $\langle N_{12}\rangle$. The normalized distribution is then defined as
 \begin{equation}
R_{12}:=\frac{N(\phi_1+\phi_2)}{\langle N_{12}\rangle} \quad.
\label{eq:r12def}
\end{equation} 
The corresponding cross section is differential in both azimuthal angles $\phi_1$,$\phi_2$ and fractional energies $z_1$,$z_2$ and thus reads \cite{daniel2}:
\begin{eqnarray}
\lefteqn{\frac{d\sigma(e^+e^-\rightarrow h_1h_2 X)}{d\Omega dz_1 dz_2 d\phi_1d\phi_2} =}& &\nonumber \\ 
&\label{eqn:sigma12} \sum_{q,\bar{q}}
\frac{3\alpha^2}{Q^2} \frac{e_q^2}{4} z_1^2 z_2^2\Bigg\{ (1+\cos^2\theta) D_1^{q,[0]}(z_1) \overline{D_1^{q,[0]}}(z_2) \quad \quad&\nonumber \\
&+  \sin^2\theta \cos(\phi_1+\phi_2) H_1^{\perp,[1],q}(z_1) \overline{H_1^{\perp,[1],q}}(z_2) \Bigg\}\quad,&  
\end{eqnarray}
where the summation runs over all quark flavors accessible at the center-of-mass energy. Antiquark fragmentation is denoted by a bar over the corresponding quark fragmentation function; the charge-conjugate term has been omitted. The fragmentation functions do not appear in the cross section directly but as the zeroth ([0]) or first ([1]) moments in the absolute value of the corresponding transverse momenta \cite{daniel3}:
\begin{equation}
F^{[n]}(z) = \int d|\mathbf{k}_T|^2 \left[\frac{|\mathbf{k}_T|}{M}\right]^n F(z,\mathbf{k}_T^2)\quad .
\end{equation}
In this equation the transverse hadron momentum has been rewritten in terms of the intrinsic transverse momentum of the process: $\mathbf{P}_{h\perp} = z \mathbf{k}_T$. The mass $M$ is usually set to be the mass of the detected hadron, in the analysis presented here $M$ will be the pion mass. 
\\

\begin{figure}[t]
\begin{center}
\includegraphics[width=7cm]{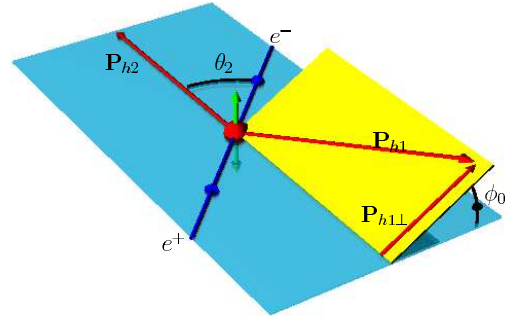}
\caption{Definition of the azimuthal angle $\phi_0$ formed between the planes defined by the lepton momenta and that of one hadron and the second hadron's transverse momentum $P_{h1\perp}'$ relative to the first hadron.}
\label{angle3}
\end{center}
\end{figure}

A second way of calculating the azimuthal asymmetries, method $M_0$, integrates over all thrust axis directions leaving only one azimuthal angle. This angle is defined as the angle between the planes spanned by one hadron momentum and the lepton momenta, and the transverse momentum of the second hadron with respect to the first hadron momentum. This angle in the opposite jet hemisphere is displayed in Fig.~\ref{angle3}, and is calculated as 
\begin{eqnarray}
\phi_0 & =& sgn \left[\mathbf{P}_{h2} \cdot \left\{ (\mathbf{\hat{z}}\times \mathbf{P}_{h2})  \times (\mathbf{P}_{h2} \times \mathbf{P}_{h1}) \right\}\right] \nonumber \\&\times &\arccos \left(\frac{\mathbf{\hat{z}}\times \mathbf{P}_{h2}}{|\mathbf{\hat{z}}\times\mathbf{P}_{h2}|}  \cdot \frac{\mathbf{P}_{h2} \times \mathbf{P}_{h1}}{|\mathbf{P}_{h2}\times \mathbf{P}_{h1}|}\right) \ . 
\end{eqnarray}
The corresponding normalized distribution $R_0$, which is defined as 
\begin{equation}
R_0:=\frac{N(2\phi_0)}{\langle N_0 \rangle} \quad,
\label{eq:r0}
\end{equation} 
contains a $\cos(2\phi_0)$ modulation.
The differential cross section depends on fractional energies $z_1,z_2$ of the two hadrons, on the angle $\phi_0$ and the transverse momentum $Q_T=|\mathbf{q}_T|$ of the virtual photon from the $e^+e^-$ annihilation process in the two hadron center-of-mass system. At leading order in $\alpha_S$ and $1/Q^2$ it assumes the form \cite{daniel}:
\begin{eqnarray}\label{eqn:sigma0}
\frac{d\sigma(e^+e^-\rightarrow h_1h_2 X)}{d\Omega dz_1 dz_2 d^2\mathbf{q}_T} =
\frac{3\alpha^2}{Q^2} z_1^2 z_2^2\Bigg\{ A(y) \mathcal{F}[D_1 \overline {D_2}] \quad + \quad\nonumber \\
B(y) \cos(2\phi_0)\mathcal{F}\left[(2\mathbf{\hat{h}}\cdot \mathbf{k}_T  \mathbf{\hat{h}}\cdot \mathbf{p}_T - \mathbf{k}_T\cdot \mathbf{p}_T ) \frac{H_1^\perp \overline{H_2^\perp}}{M_1M_2}\right] \Bigg\},  
\label{convolution}
\end{eqnarray}
where the convolution integral over the transverse momenta $\mathbf{k}_T=z_1\mathbf{P}_{h1\perp}$ and $\mathbf{p}_T=z_2\mathbf{P}_{h2\perp}$ is abbreviated as:
\begin{equation}
\mathcal{F}[X] = \sum_{a,\bar{a}} e_a^2 \int d^2\mathbf{k}_T d^2\mathbf{p}_T \delta^2(\mathbf{p}_T + \mathbf{k}_T - \mathbf{q}_T) X \quad.
\end{equation}
$\hat{\mathbf{h}} $ denotes a unit vector in the direction of the transverse momentum of the first hadron relative to the axis defined by the second hadron.
In Eqn.~(\ref{convolution}) as well as in some formulas to follow, the explicit z dependence of the fragmentation functions will be omitted while the indices will be partially retained: $D_1^q(z_1) \rightarrow D_1$, $\overline{D}^q_1(z_2)\rightarrow\overline{D}_2$ and similarly for the Collins functions $H_1^\perp$ and $\overline{H}_2^\perp$. 
The kinematic factors are defined as:
\begin{eqnarray}
A(y) = & (\frac{1}{2} -y - y^2) &\stackrel{CMS}{=} \frac{1}{4} (1+ \cos^2\theta) \\
B(y) = & y(1 - y) &\stackrel{CMS}{=} \frac{1}{4} (\sin^2\theta)\quad,
\end{eqnarray}
where $y=(1+\cos\theta)/2$ is a measure of the forwardness of the scattering process.
The cross sections given in (\ref{eqn:sigma12}) and (\ref{eqn:sigma0}) are related. Integrating either over the azimuthal angles $\phi_1$ and $\phi_2$ or over the azimuthal angle $\phi_0$ and $q_T$ in the other case will give the same unpolarized cross section. Similarly, the Collins contributions can be related to each other, but here, due to the additional convolutions of transverse momenta, one either has to know the intrinsic transverse momentum dependence of the Collins function or rely on assumptions. A majority of authors assume that the Collins function is a Gaussian in $k_t$ but of different width than the unpolarized fragmentation function \cite{alessandro,alexei,leonard,peter}. Measuring the two types of asymmetries one can infer the intrinsic transverse momentum dependence of the Collins function. The calculation of the asymmetries using a Gaussian $k_t$ distribution will be discussed further in section \ref{sec:param}.
\subsection{Unlike-sign, like-sign and charged pion pairs}
Favored fragmentation functions describe the fragmentation of a quark of flavor q into a hadron with a valence quark of the same flavor. For example, we refer to the fragmentation processes $u\rightarrow \pi^+$ and  $d\rightarrow \pi^-$ (and charge-conjugates) as favored fragmentation and to the fragmentation processes  $u\rightarrow \pi^-$ and $d\rightarrow \pi^+$ (and charge-conjugates) as disfavored fragmentation processes. 
We define yields, $N^U$ for unlike-sign inclusive pion pair production: $e^+e^- \longrightarrow \pi^\pm\pi^\mp X$; $N^L$ for like-sign pion production $e^+e^- \longrightarrow \pi^{\pm}\pi^{\pm}X$ and $N^C$ for charged pions without charge sign identification: $e^+e^- \rightarrow \pi \pi X$. Consider for example, the production of unlike-sign charged pions from a pair of up- and anti-up-quarks: $e^+e^- \rightarrow u\bar{u} \rightarrow \pi^{\pm}\pi^{\mp}X$. The pion pair can be either created through two favored fragmentation processes, $D_u^{\pi^+}(z_1) \times D_{\bar{u}}^{\pi^-}(z_2)$ or, suppressed in yield, through two disfavored fragmentation processes: $D_u^{\pi^-}(z_1) \times D_{\bar{u}}^{\pi^+}(z_2)$.
 We introduce the favored fragmentation functions $ D^{fav}_{1} = D^{fav}(z_{1})= D_u^{\pi^+}(z_1)$ and $ \bar{D}^{fav}_{2} = D^{fav}(z_{2})= D_{\bar{u}}^{\pi^-}(z_2)$ as well as the disfavored fragmentation functions $ D^{dis}_{1} = D^{dis}(z_{1})= D_u^{\pi^-}(z_1)$ and $ \bar{D}^{dis}_{2} = D^{dis}(z_{2})= D_{\bar{u}}^{\pi^+}(z_2)$.
Ignoring strange- and heavy quark fragmentation into pions and assuming SU(2) flavor symmetry the cross sections for charged pion pair production can be written as:
\begin{eqnarray}
N^U&=&\frac{d\sigma(e^+e^-\rightarrow \pi^\pm\pi^\mp X)}{d\Omega dz_1 dz_2}   \approx\frac{\alpha^2}{3Q^2} (1+\cos^2\theta)\nonumber \\  
&& \sum_q e_q^2  \left(D_1^{fav} \overline {D_2}^{fav} + D_1^{dis} \overline {D_2}^{dis}\right) \\
N^L&=&\frac{d\sigma(e^+e^-\rightarrow \pi^\pm\pi^\pm X)}{d\Omega dz_1 dz_2}  \approx \frac{\alpha^2}{3Q^2}(1+\cos^2\theta)\nonumber \\
&& \sum_q e_q^2\left(D_1^{fav} \overline {D_2}^{dis} + D_1^{dis} \overline {D_2}^{fav}\right)  \\
N^C&=&\frac{d\sigma(e^+e^-\rightarrow \pi\pi X)}{d\Omega dz_1 dz_2}\approx\frac{\alpha^2}{3Q^2}(1+\cos^2\theta)\nonumber \\
&& \sum_q e_q^2\left(D_1^{fav}+ D_1^{dis}\right)  \left(\overline {D_2}^{fav}+ \overline {D_2}^{dis}\right).  
\end{eqnarray}
\subsection{Competing sources for azimuthal correlations}
In addition to the Collins fragmentation of back-to-back pion pairs there are other processes that also can result in azimuthal correlations between the pions. A detailed understanding of these backgrounds is crucial for the extraction of the Collins function.  
\subsubsection{$\gamma-Z$ interference}
Pure Z exchange at Belle energies can be safely ignored. We briefly consider contributions from $\gamma Z$ interference. Flavor dependent changes to the coupling $e_q^2/4$ amount to corrections to the asymmetries of $1.0004$ for $u$ quarks and $1.001$ for $d$ quarks \cite{daniel}. These corrections are small compared to the experimental sensitivity and are thus neglected.
\subsubsection{Weak decays} 
Parity violation in weak decays can lead to azimuthal correlations between hadrons in opposite hemispheres. A well known example is the $\tau$ decay channel $\tau\rightarrow \pi \nu$. We have used this channel as a test of our extraction procedure for azimuthal Collins asymmetries. We have measured azimuthal asymmetries in $\tau$ production and find results consistent with existing data \cite{aleph,delphi} and results from Monte Carlo. 

In order to remove contributions of $\tau$ leptons, which are characterized by large missing energy, from the quark-antiquark data sample we require the energy observed in the detector to be large, $E_{\rm vis}> 7$ GeV. The effect of $\tau$ background remaining after the visible energy cut was studied and has been included in the systematic error. Similarly, the production of charmed mesons with subsequent weak decays can lead to azimuthal correlations, which have been carefully studied and have been subtracted in the extraction of Collins asymmetries (see section \ref{sec:charm}).
\subsubsection{Gluon radiative effects}
The dominant Collins-like background contribution originates from low energetic gluon radiation $e^+e^-\rightarrow q\bar{q} g$, which does not manifest itself in a third jet but creates azimuthal asymmetries. The gluon radiation results in a $\cos(2\phi_0)$ modulation and according to calculations by Boer \cite{daniel3} is described by:
\begin{equation}\frac{dN}{d\Omega d^2\mathbf{q}_T} = \frac{3}{16\pi} \left[\frac{1}{2} \frac{\mathbf{q}_T^2}{Q^2 + \mathbf{q}_T^2}\sin^2\theta\cos(2\phi_0)\right]
\quad.
\end{equation}
The dependence of the asymmetry on $\mathbf{q}_T$ is nearly quadratic for small $\mathbf{q}_T$ and does not depend on the charges of the created hadrons. The differential cross section in $z_1$ and $z_2$ is proportional to the unpolarized fragmentation functions $D_1(z)$ and $\overline{D_1(z)}$ for a given charge combination of the hadrons. 
\section{The Belle experiment}
\begin{figure}[t]
\begin{center}
\includegraphics[width=8cm]{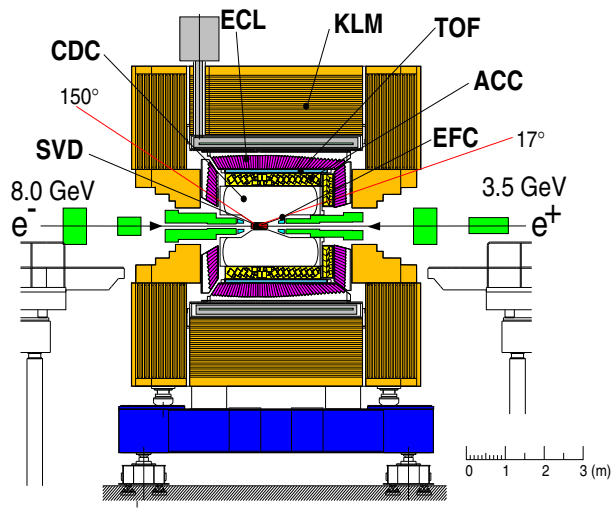}
\caption{\label{fig:belledet}Side view of the Belle detector.}
\end{center}
\end{figure}

The data were taken at the asymmetric $e^+e^-$ KEKB \cite{kekb} storage rings, which collide 8 GeV electrons and 3.5 GeV positrons. Taking into account the resulting boost between the laboratory and CMS frames the Belle detector itself is asymmetric. It is a large-solid-angle magnetic spectrometer that
consists of a multi-layer silicon vertex detector (SVD), a 50-layer central drift chamber (CDC), an array of aerogel threshold Cherenkov counters (ACC), a barrel-like arrangement of time-of-flight scintillation counters (TOF), and an electromagnetic calorimeter (ECL) comprised of CsI(Tl) crystals located inside a superconducting solenoid coil that provides a $1.5$ T magnetic field, see Fig.~\ref{fig:belledet}. An iron flux-return located outside the coil is instrumented with resistive plate chambers to detect $K^0_L$ mesons and to identify muons (KLM). The detector is described in detail elsewhere \cite{belledetector}. Two different inner detector configurations were used. For the first data sample of 155.6 fb$^{-1}$, a $2.0$ cm radius beam pipe and a 3-layer silicon vertex detector were used; the rest of the 547 fb$^{-1}$ of data has been collected using a $1.5$ cm radius beam pipe surrounded by a 4-layer silicon detector and a small-cell inner drift chamber \cite{Belle2}.
Particle identification in Belle is done by combining data from the TOF, ACC and CDC-dE/dx subdetectors.
\subsection{Monte Carlo}
In order to correct the data for detector effects and for systematic studies MC simulated events were generated by the Pythia 6.2 or qq98 \cite{jetset} generators and processed with a full simulation of the Belle detector based on the GEANT3 package \cite{geant}. The MC events are processed with the same reconstruction algorithms and written to the same data structure as real data. The MC software package has a built-in backtracing facility for final state particles in order to facilitate the assessment of detector effects.
In the center-of-mass system the Pythia program generates a back-to-back quark-antiquark pair with a $1+\cos^2\theta$ distribution relative to the $e^+e^-$ beam axis. The next step is a perturbative parton cascade based on order $\alpha_s$ matrix elements in the leading log approximation, which is followed by a parton shower fragmenting into the final state hadrons. 
Within the parton shower and hadronization processes transverse momenta are being generated, which could lead to azimuthal asymmetries. The Collins effect is not modeled in the standard event generators used for MC production in Belle. 
The MC samples are divided according to the type of event generated: charged and neutral {\it B} meson pairs created from the $\Upsilon(4S)$, light quark (uds) pair production, charm quark pair production, and $\tau^+\tau^-$ production (using the Tauola package \cite{tauola}). In what follows, MC refers to the light quark MC simulation if not specified otherwise.
\section{Analysis}
We report results obtained with an integrated luminosity of 547 fb$^{-1}$. A 55 fb$^{-1}$ sample was taken at CM energy of $10.52$ GeV while 492 fb$^{-1}$ was accumulated on the $\Upsilon(4S)$ resonance at $10.58$ GeV. 
At the lower CMS energy, which is below the threshold for $B\bar{B}$ meson pair production, only light and charm quark pair creation contribute to the hadronic final states. In the higher energy data in addition to continuum events there are resonant $\Upsilon(4S)$ decays into neutral and charged {\it B} meson pairs.
\subsection{Event and track selection\label{sec:eventsel}}
\begin{figure}[t]
\begin{center}
\includegraphics[width=7cm]{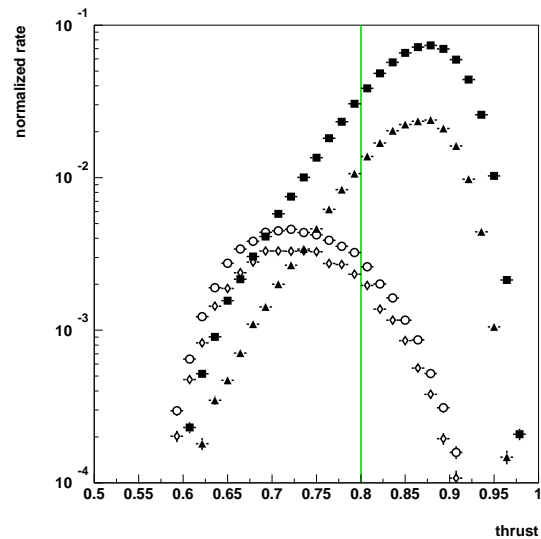}
\caption{\label{fig:thrust} Simulated thrust distributions for selected 2-pion pairs at $\sqrt{s}=10.58$ GeV, for $e^+e^-\rightarrow B^+B^-$ events (open diamonds), $e^+e^-\rightarrow B^0\bar{B}^0$ events (open circles), $e^+e^-\rightarrow c\bar{c}$ events (full triangles) and for light quark production $e^+e^-\rightarrow q\bar{q},q\in uds$ (full squares) normalized to the total number of events in all channels. The vertical line represents the minimal thrust value selected for the analysis.} 
\end{center}
\end{figure}

The Collins effect is expected to be dominant in the fragmentation of light quarks as helicity is only conserved for nearly massless quarks while for heavier quarks the correlation between the quark and the antiquark side may be lost. We also focus on the measurement of the Collins effect in light quark fragmentation, as it is the light quark Collins fragmentation function that is needed as input for studies of transverse proton spin structure in semi-inclusive deep inelastic scattering or polarized proton-proton collisions. 
Most of the {\it B} meson events can be removed from the data sample, using the difference in event shapes between events with underlying {\it B} mesons and light quarks.
Since the {\it B} mesons decay nearly at rest in the CMS, the final state particles exhibit a more spherical spatial distribution, which corresponds to low thrust values. Most of the light quark-antiquark pairs appear in a two-jet topology, which corresponds to high thrust values. Consequently, for pion pairs a thrust cut of $T>0.8$ removes 98\% of B data as can be seen in Fig.~\ref{fig:thrust}, where  the simulated thrust distributions for light and charmed quark pairs and $\Upsilon(4S)$ decays are shown. 

For the calculation of the thrust variable all charged tracks and all neutral particles with a minimum energy of $0.1$ GeV are considered. For the purpose of obtaining an unbiased data sample one assigns the sign of the thrust axis at random.
The contribution from $B$ mesons to the observed asymmetries can be estimated by comparing the data taken on the $\Upsilon(4S)$ resonance with the data taken 60 MeV below the resonance. This test will be discussed in Section \ref{sec:consistency}.
Events with charm quarks do not exhibit a very different event shape from light quark events, see Fig.~\ref{fig:thrust}. However, the contributions from events with charm quarks can be corrected by measuring azimuthal asymmetries in a charm-enhanced data sample. This will be described in Section \ref{sec:charm}.

In order to ensure a two-jet geometry in the selected event sample with the majority of final state particles reconstructed in one of the two jets a minimum visible energy of $E_{visible}>7$\, GeV is required. 
Charged tracks used in the analysis are required to originate from the interaction point and to lie in a fiducial region $-0.6 <  \cos(\theta_{\rm lab}) < 0.9$, where $\theta_{\rm lab}$ is the polar angle in the laboratory frame relative to the direction opposite the the incoming positron (definition of the $z$-axis). This corresponds to a nearly symmetric fiducial interval in the CMS frame $-0.79 < \cos \theta_{\rm CMS} < 0.74 $ and covers the acceptance of the barrel part of the Belle detector. For the identification of pions a likelihood ratio is used, which is based on energy loss in the drift chamber (CDC), the number of Cherekov photons (ACC) and time of flight information (TOF). Kaons are separated from pions by requiring $\mathcal{L}(\pi)/[\mathcal{L}(K) + \mathcal{L}(\pi)] > 0.7 $. $\mathcal{L}(\pi/K)$ is the likelihood for a track be a pion or kaon. The percentage of misidentified pion pairs is below 10\% in all $z_1$ and $z_2$ bins. In addition, the likelihood ratios for being either a muon or an electron have to be below 0.9 and 0.8, respectively. A cut on the fractional hadron energy of the two hadrons $z_{1,2}= 2E_{1,2}/Q>0.2$, avoids contributions from decays with the decay products incorrectly reconstructed in opposite hemispheres.

\begin{figure}[t]
\begin{center}
\includegraphics[width=7cm]{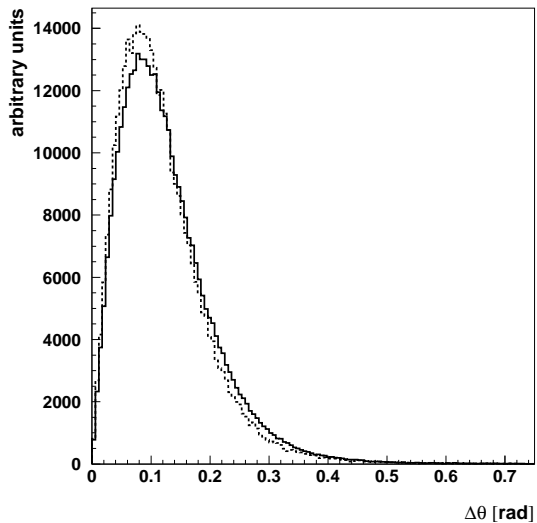}
\caption{\label{fig:deltathrust} Angular deviation $\Delta \theta$ from the original quark-antiquark axis of the reconstructed thrust axis (solid line) and of the thrust axis calculated from generated events before detector simulation (dashed line). The distributions are based on a light quark MC simulation with a $T>0.8$ requirement for the generated events.} 
\end{center}
\end{figure}

The two pion tracks are required to lie in opposite hemispheres with the selection $W_{\rm hemi}:=(\mathbf{P}_{h1} \cdot \mathbf{\hat{n}})(\mathbf{P}_{h2} \cdot \mathbf{\hat{n}}) < 0 $, where the hemispheres are separated by the plane normal to the thrust axis $\mathbf{\hat{n}}$.  A comparison of the quark-antiquark axis with the thrust axis calculations from reconstructed particles shows an average angular deviation between the two of $128\pm 82$ mrad (the RMS value is quoted for the uncertainty) in simulated events for light quark production, while it appears to be slightly larger for charm production (see Fig.~\ref{fig:deltathrust}) due to semileptonic decays. Since the thrust axis calculated from generated particles also deviates from the original quark-antiquark axis by a similar magnitude, we conclude that the observed deviation arises from the intrinsic difference between the original quark direction and thrust axis in the hadronization process. 
A test with the approximate algorithm buildt in Pythia confirms these deviations. The smearing of the thrust axis compared to the quark axis leads to a reduction in the amplitudes of the measured azimuthal back-to-back correlations, as discussed below.

The impact of initial state radiation (ISR) from the incoming leptons was estimated using MC simulation. The average CMS energy for events fulfilling all selection criteria is reduced by ISR to 10.51 GeV for the on-resonance data with about 2\% of events at energies below 9 GeV. Hence the fractional energies are slightly underestimated, but the effect of ISR is negligible compared to the width of the $z$-bins used in this analysis. The effect of ISR on the asymmetry measurements was inferred from a MC simulation employing a reweighting technique, which is discussed in section \ref{sec:reweighting}. The results were consistent with those obtained without ISR and thus, the overall effect of QED initial state radiation on this analysis is negligible. 

\begin{figure}[t]
\begin{center}
\includegraphics[width=7cm]{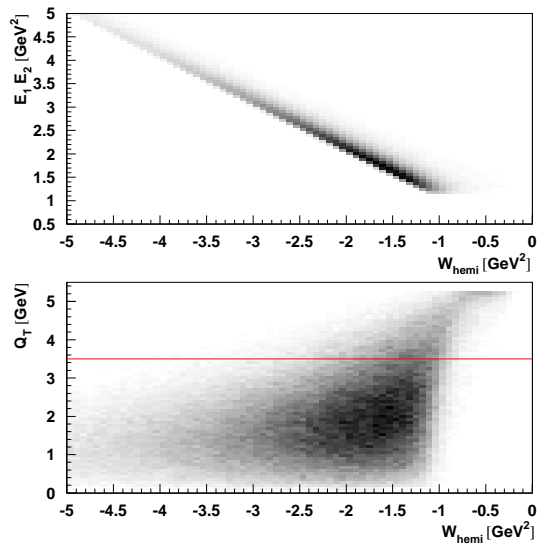}
\caption{\label{fig:hemi}Hemisphere variable $W_{\rm hemi}$ compared to the product of energies of the two hadrons (top) and the transverse virtual photon momentum $Q_T$ (bottom plot) for pion pairs with $W_{\rm hemi}>-5\ \mathrm{GeV}^2$. The horizontal line corresponds to the maximum $Q_T$ value selected for the analysis.} 
\end{center}
\end{figure}

 Possible hemisphere misassignment is further reduced by the requirement that the transverse momentum of the virtual photon in the two-pion CMS ($Q_T$) be smaller than 3.5 GeV. The high $Q_T$ region is mainly populated by pion pairs with very asymmetric fractional energies $z$ and especially when the lower energetic particle happens to be close to the hemisphere boundary (see Fig.~\ref{fig:hemi}). For such pairs the remaining uncertainty in the determination of the thrust axis direction can lead to a hemisphere misassignment and the simple correlation of one pion originating from the quark side and the other pion originating from the antiquark side will be spoiled. At the hemisphere boundary asymmetric particle decays could also lead to the assignment of one pion into the wrong hemisphere. 
After application of the $Q_T$ cut the hemisphere is misassigned in less than 0.1\% of the events according to MC simulation.
\subsection{Normalized yields, raw asymmetries}
In the expression for the di-hadron yields in (\ref{eqn:sigma0}) and (\ref{eqn:sigma12}) the product of two Collins functions appears as the amplitude of a cosine modulation in the di-hadron yield. The Collins functions depend on the fractional energies $z_{1,2}$ of the two hadrons.
Therefore the analysis is performed in 4 bins in $z_1$ and $z_2$ for each hadron with boundaries at $z_i = 0.2$, $0.3$, $0.5$, $0.7$ and $1.0$. Alternatively, the cosine modulation will be evaluated in bins of $Q_T$. In this case the dependence on the fractional energies is averaged over all bins in $z_1$ and $_2$. 
As a first step in the analysis, the two-pion yields $N_\alpha(\beta_\alpha)$ are obtained for each $(z_1,z_2)$ bin in 8 equidistant bins of the azimuthal angles $\beta_\alpha, \alpha=0,12$. The index $\alpha=12$ refers to the thrust method $M_{12}$ introduced in (\ref{eqn:sigma12}) and the angle $\beta_{12} =  \phi_1+\phi_2$ is the sum of the hadron angles shown in Fig.~\ref{angle2}. The index $\alpha=0$ refers to the two hadron method $M_0$ as introduced in Eqn.~(\ref{eqn:sigma0}) and the angle $\beta_0=2\phi_0$ is defined in Fig.~\ref{angle3}.
Next, the azimuthal di-hadron yields, $N_\alpha(\beta_\alpha)$ are normalized to the average di-hadron yield $\langle N_\alpha \rangle$ per bin resulting in the normalized yields $R_\alpha$, defined in Eqns.~(\ref{eq:r12def}) and (\ref{eq:r0}). 
Following Eqns.~(\ref{eqn:sigma0}) and (\ref{eqn:sigma12}) the normalized yields can be parameterized as:
\begin{eqnarray}
R_{\alpha}& =& a_{\alpha} \cos(\beta_\alpha) + b_{\alpha}, (\alpha = 0,12). 
\label{eq:yieldfit}
\end{eqnarray}
\begin{figure}[t]
\includegraphics[width=7cm]{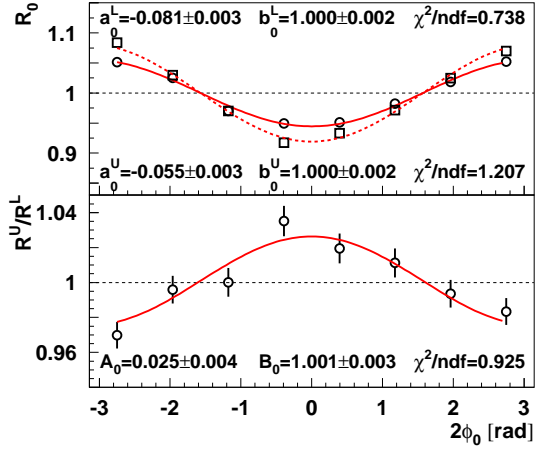}% Here is how to import EPS art
\caption{\label{fig:rawfit}Raw data. Top: Example of uncorrected unlike-sign (open circles) and like-sign (open squares) pion pair normalized rate $R_0$ vs.~$2\phi_0$ in the bin $z_1 (z_2) \in
[0.5,0.7]$, $z_2 (z_1) \in [0.3,0.5]$. Bottom: Pion pair
double ratio $R_0^U/R_0^L$ vs.~$2\phi_0$ in the same $z_1,z_2$ bin. The results of the fit described in the text (full and dashed lines) are also shown.}
\end{figure}
\begin{figure}[t]
\begin{center}
\includegraphics[width=8.5cm]{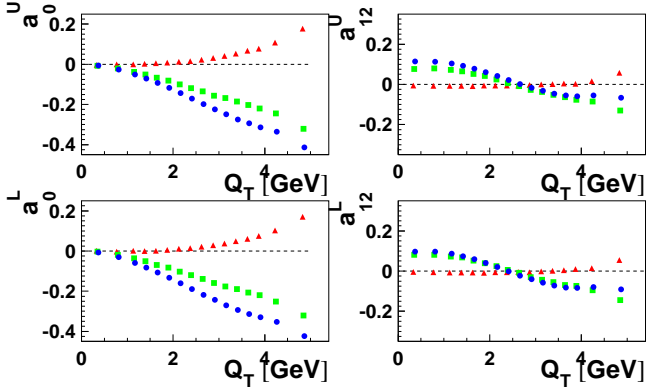}
\caption{\label{fig:aulraw} Raw asymmetry parameters $a_\alpha^U$ for unlike-sign pion pairs and $a_\alpha^L$ for like-sign pairs as a function of the transverse virtual photon momentum $Q_T$. Light quark MC generated (triangles), reconstructed (squares), and selected data sample (circles) for the $\alpha=0,12$ reconstruction methods are shown.}
\end{center}
\end{figure}
The measured normalized yields $R_\alpha$ are fitted with the parameterization in (\ref{eq:yieldfit}) leaving $a_\alpha$ and $b_\alpha$ as free parameters. The constant terms $b_\alpha$ are found to be consistent with unity within the statistical errors as shown in Fig.~\ref{fig:rawfit}. The raw asymmetry parameters, $a_\alpha$, are found to be large as can be seen in Fig.~\ref{fig:aulraw} for the unlike-sign (U) and like-sign (L) pion pairs as a function of $Q_T$. The figure compares asymmetries for pion pairs from data with asymmetries for pion pairs from generated and reconstructed MC data samples. All track and event selection cuts have been applied with the exception of the cut on the transverse photon momentum $Q_T<3.5$\, GeV. No particle identification and polar angle selection cuts have been applied for the generated MC sample. 
\subsection{Double ratios}
The $a_0$ raw asymmetries obtained from generated MC events are nonzero and increasing with $Q_T$ while the $a_{12}$ are almost zero. A large difference between the asymmetries calculated directly from generated MC events and the asymmetries obtained from MC events after GEANT simulation of the detector response and track reconstruction can be observed in Fig.~\ref{fig:aulraw}. This difference points to large acceptance effects, which can even cause the $A_0$ asymmetries to change sign in method $M_0$.
We have studied the possibility of cancelling acceptance effects in the asymmetries $a_0$ and $a_{12}$ using a MC simulation of the acceptance effects but found it difficult to separate asymmetries from the Collins effect from acceptance effects and the radiative background in $a_\alpha$.
The small differences between the asymmetries $a_\alpha$ from reconstructed MC events and real data gives an estimate on how well radiative effects are described in the MC event generator.
Therefore in order to correct the raw asymmetries for detector effects one has to rely heavily on a MC simulation, which does not include any spin-dependent asymmetries. As a consequence we do not consider these any further and employ a scheme using ratios of asymmetries, in which most of the instrumental effects cancel, and is thus independent of the details of the MC simulation.
\subsubsection{Method}
The asymmetries generated by QCD radiative events and acceptance effects do not depend on the charge combination of the pion pairs. For detector effects this can be tested by comparing positively charged with negatively charged pion pairs, as will be discussed in section \ref{sec:chargedratios}. 
When building the normalized yields $R_\alpha$ the normalization is obtained by integrating over the azimuthal angles $\beta_\alpha$. This causes the terms proportional to $\cos(\beta_\alpha)$ to vanish. Adding to the cross section in (\ref{eqn:sigma12}) the term due to gluon radiation, which is proportional to the unpolarized fragmentation functions, and has been scaled by a factor C, and dividing by the average number of hadron pairs in the full $\beta_\alpha$ range, we obtain expression (\ref{eq:r12}) for $R_{12}$.
\begin{widetext}
 \begin{eqnarray}
R_{12}&=&\frac{N(\phi_1+\phi_2)}{\langle N_{12}\rangle } \nonumber \\ & \propto&\Bigg[(1+\cos^2\theta)\sum_q e_q^2 D_1(z_1) \overline{D}_1(z_2) +  \sin^2\theta\cos(\phi_1+\phi_2)\Big[\sum_q e_q^2 f(H_1^\perp(z_1)\overline{H}_1^\perp(z_2))  \nonumber \\&+& C \sum_q e_q^2 D_1(z_1) \overline{D}_1(z_2)\Big]\Bigg] \cdot \left[(1+\cos^2\theta)\sum_q e_q^2 D_1(z_1) \overline{D}_1(z_2)\right]^{-1}  \nonumber \\&=& 1 + \frac{\sin^2}{1+\cos^2\theta}\cos(\phi_1+\phi_2)\left[\frac{\sum_q e_q^2 f(H_1^\perp(z_1)\overline{H}_1^\perp(z_2))}{\sum_q e_q^2 D_1(z_1) \overline{D}_1(z_2)} + C\right]\ .\label{eq:r12}
\end{eqnarray} 
\end{widetext}
In order to obtain a MC independent measure of the Collins effect, one can form a double ratio of, for example, the normalized yields of unlike-sign over like-sign pion pairs:
\begin{eqnarray}
  \frac{R^U_{\alpha}}{R^L_{\alpha}}&:=&\frac{N^U_{\alpha}(\beta_\alpha)/\langle N^U_{\alpha} \rangle}{N^L_{\alpha}(\beta_\alpha)/\langle N^L_{\alpha} \rangle} \ , (\alpha=0,12).
\end{eqnarray}
Up to the linear term in the amplitude of $\cos(\beta_\alpha)$ the term multiplied by C in (\ref{eq:r12}) cancels and one is left with the following expression for the double ratio:
\begin{widetext}
\begin{eqnarray}
\frac{R_{12}^U}{R_{12}^L}&=&1+ \cos(\phi_1+\phi_2) \frac{\sin^2\theta}{1+cos^2\theta}%\nonumber \\ 
\Bigg\{\frac{f\left(H_1^{\perp,fav}\overline{H}_2^{\perp,fav} + H_1^{\perp,dis}\overline{H}_2^{\perp,dis}\right)}{ \left(D_1^{fav}\overline{D}_2^{fav} + D_1^{dis}\overline{D}_2^{dis}\right) } -%\nonumber \\ 
\frac{f\left(H_1^{\perp,fav}\overline{H}_2^{\perp,dis}\right)}{ \left(D_1^{fav}\overline{D}_2^{dis}\right)}\Bigg\}.
\label{eq:dr}
\end{eqnarray}
\end{widetext}
Similarly one obtains the double ratio of unlike-sign and any sign (i.e. $++$, $+-$ and charge-conjugates; C) pion pairs:
\begin{widetext}
\begin{eqnarray}
\frac{R_{12}^U}{R_{12}^C}& = & 1+ \cos(\phi_1+\phi_2) \frac{\sin^2\theta}{1+\cos^2\theta} \nonumber \\
& \times & \Bigg\{\frac{f\left(H_1^{\perp,fav}\overline{H}_2^{\perp,fav} + H_1^{\perp,dis}\overline{H}_2^{\perp,dis}\right)}{ \left(D_1^{fav}\overline{D}_2^{fav} + D_1^{dis}\overline{D}_2^{dis}\right) } \frac{f\left((H_1^{\perp,fav} + H_1^{\perp,dis})(\overline{H}_2^{\perp,fav}+\overline{H}_2^{\perp,dis})\right)}{ \left((D_1^{fav}+D_1^{dis})(\overline{D}_2^{fav}+\overline{D}_2^{dis})\right)}\Bigg\}.
%\frac{R_{12}^U}{R_{12}^C}=1+ \cos(\phi_1+\phi_2) \frac{\sin^2\theta}{1+\cos^2\theta}&
%\Bigg\{\frac{f\left(H_1^{\perp,fav}\overline{H}_2^{\perp,fav} + H_1^{\perp,dis}\overline{H}_2^{\perp,dis}\right)}{ \left(D_1^{fav}\overline{D}_2^{fav} + D_1^{dis}\overline{D}_2^{dis}\right) }  & \nonumber \\ 
%- &\frac{f\left((H_1^{\perp,fav} + H_1^{\perp,dis})(\overline{H}_2^{\perp,fav}+\overline{H}_2^{\perp,dis})\right)}{ \left((D_1^{fav}+D_1^{dis})(\overline{D}_2^{fav}+\overline{D}_2^{dis})\right)}\Bigg\}.&
\label{eq:dr2}
 \end{eqnarray}
\end{widetext}
This double ratio contains a different combination of favored and disfavored fragmentation functions as pointed out by Efremov and Schweitzer \cite{peter}.
Analogous expressions can be obtained for the $R^i_{0}/R^j_{0}$ double ratios with $i,j \in \mathrm{U,L,C}$. In the double ratios the acceptance effects cancel while the QCD radiative effects cancel to first order. The possible influence of higher order terms ($\cos^2(\beta_\alpha)$) has been studied explicitly by including these terms in the fits and also by performing a subtraction of U and L (C) normalized yields, as discussed in the next section. Statistical correlations between the U, L and C pairs have been taken into account.
The double ratios are parameterized by $R^{i}_\alpha /R^{j}_\alpha = A^{ij}_\alpha \cos(\beta_\alpha) + B^{ij}_\alpha$, and the measured distributions are fitted with $A_\alpha^{ij}$ and $B_\alpha^{ij}$ as free parameters. The constant terms $B_\alpha^{ij}$ are again found to be consistent with unity within statistical errors.
\section{Systematic studies\label{sec:systematics}}
The impact of the detector performance and of the method used for asymmetry reconstruction on the isolation of the Collins effect has been estimated through various systematic studies.
\subsection{MC double ratios}
\begin{figure}[th]
\begin{center}
\includegraphics[width=7cm]{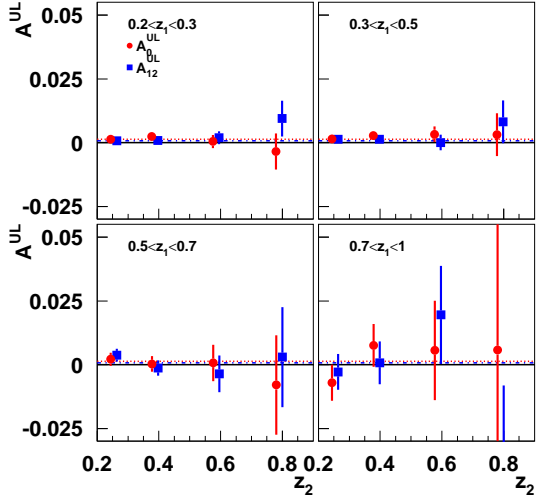}
\caption{Asymmetry parameters $A^{UL}_0$(circles) and $A^{UL}_{12}$(squares) as a function of $z_2$ for 4 $z_1$ bins obtained from MC double ratios. The results of fits to a constant are also displayed as dash-dotted lines. They are $(0.13\pm0.09)\%$ for $A_0^{UL}$, $(0.08\pm0.09)\%$ for $A_{12}^{UL}$, $(0.06\pm0.04)\%$ for $A_0^{UC}$ and $(0.04\pm0.04)\%$ for $A_{12}^{UC}$. The UC data points are not shown.} 
\label{fig:mc}
\end{center}
\end{figure}

An important test of the double ratio method is the extraction of double ratios from MC. The generic MC describes the radiative gluon effects. The acceptance effects are also included in the detector simulation. However, the MC generator does not contain asymmetries based on the Collins effect. Therefore one expects the cosine moments of the double ratios for MC to vanish. Figure \ref{fig:mc} shows the fitted asymmetry parameters $A^{UL}_\alpha$ as a function of the $z_1,z_2$ bins. It can be seen that these asymmetries are consistent with zero in all bins. Similar features are observed for the asymmetry $A^{UC}$, which is not shown. A fit of a constant to the observed asymmetries $A_\alpha^{UL}$ and $A_\alpha^{UC}$ was performed and the largest deviations from zero, including the statistical errors of the fits, were attributed as systematic uncertainties of the final results. The contribution to the systematic uncertainty ranges from 0.07\% in $A^{UC}_{12}$ to 0.22\% in $A^{UL}_{0}$ as summarized in Table \ref{tab:systematics}.  
\subsection{Single spin asymmetries}
The Collins fragmentation in $e^+e^-$ annihilation applies when the two primary quarks have equal but opposite transverse spin components; asymmetries only occur when hadrons from both quarks are observed simultaneously. For unpolarized beams there is no net transverse polarization of an individual quark or antiquark, and hence there is no modulation of $\phi_1$ or $\phi_2$ separately. Using the same event selection, and measuring $\sin\phi_1 $ or $\sin\phi_2$ modulations we found these asymmetries to be consistent with zero within the statistical uncertainty. No systematic error is assumed.  
\subsection{Event mixed asymmetries}
The spin information of quark-antiquark pairs also is lost if one uses the same analysis procedure but combines two pions from different events. Such pion pairs would be uncorrelated and hence no asymmetries should be visible. This test has been performed and the results were consistent with zero asymmetries with the largest asymmetries being below 0.01\%. No contribution to the systematic error has been assigned. 

\subsection{Moment reconstruction in reweighted simulated samples\label{sec:reweighting}}
\begin{figure}[t]
\begin{center}
\includegraphics[width=7cm]{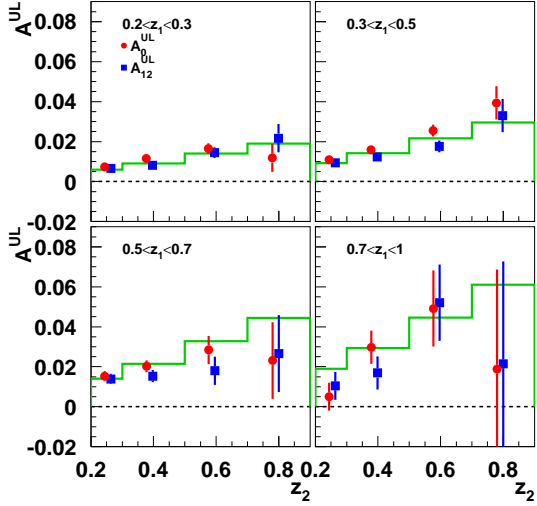}
\caption{Double ratio results $A^{UL}_{0}$ (circles) and $A^{UL}_{12}$ (squares) as a function of $z_2$ for 4 $z_1$ bins for reconstructed light quark MC, reweighted as described in the text. The generated asymmetry $0.1z_1z_2$ evaluated at the bin centers is also shown as histogram. The dotted line corresponds to the zero asymmetry.}
\label{fig:zweighted}
\end{center}
\end{figure}

The MC generators do not include spin effects, therefore in order to perform a realistic simulation, which is as close as possible to the experimental observations, weights have been introduced to generate asymmetries. This allows to study the influence of the detector effects on the asymmetries. The weights $w^i$ are applied to the generated particles before detector simulation and depend on the azimuthal angles $\beta_\alpha$ and the fractional energies $z_{gen}$ of the final state particles.
\begin{equation}
w^i= 1 + a^i(z_{gen}) \cos{\beta_{\alpha,gen}}\ , (i=U,L,C;\alpha=0,12)\quad.
\end{equation}
Analyzing the UL double ratios with, for example, generated weights $w^U=5\%$ and $w^L=-5\%$, the reconstructed double ratios should ideally return a 10\% asymmetry. 
An example of the reconstructed UL asymmetries as a function of the $z$ variables is shown in Fig.~\ref{fig:zweighted}, compared to the linear weights in $z_1$ and $z_2$ of $w^U=1+0.05\cdot z_1z_2\cos(\beta_\alpha)$ and $w^L=1-0.05\cdot z_1z_2\cos(\beta_\alpha)$. A summary of several combinations of $z_{1,2}$ independent weights and their reconstructed values (as obtained by a constant fit to all $z$ bins) are displayed in Table \ref{tab:weighted}.
The $A_0$ results are consistent with the generated asymmetries, while the $A_{12}$ results systematically underestimate the generated asymmetries. This underestimation can be attributed to the difference between the reconstructed thrust axis and the original quark-antiquark axis. The direction of this axis is essential in the estimation of  $A_{12}$; because of the deviation the calculated azimuthal angles and also their modulations are smeared and thus the amplitudes of the modulations decrease. An average underestimation of $ (58.9\pm1.3)$  %$17.6\pm 1.1$
\% in the $A_{12}$ results is observed. In order to obtain asymmetries relative to the quark-antiquark axis these results are thus rescaled by  a factor $1.66\pm 0.04$%$1.21\pm0.01$
. The corrected asymmetries can now be used at all energies, because the deviation between the thrust axis and the quark directions, which might be energy dependent, is removed.
 The $A_0$ asymmetries depend only marginally on the original quark-antiquark axis due to the hemisphere assignment. The small underestimation of the asymmetries can thus be attributed to the smearing of the tracks in the detector. To correct for this dilution, the $A_0$ results are rescaled by a factor of $ 1.11\pm0.05$. The error on the correction factor is added as systematic error (see Table \ref{tab:systematics}).

\begin{table}[h]
\caption{\label{tab:weighted}$z$-independent weights generated in the $U$, $L$ and $C$ channels using generated tracks and the original quark-antiquark axis and the reconstructed average asymmetries as obtained by a constant fit to all $z$ bins.}
%\begin{tabular}{ccccc}\hline
%\multicolumn{3}{c}{weights}&\multicolumn{2}{c}{UL double ratios}\\
%$a_U$ & $a_L$& $a_C$& $A_0$& $A_{12}$\\ \hline
%+5 &-5& -5 &$(9.8 \pm 0.1)\%$&$ (5.9 \pm 0.2)\%$\\ 
%10 &0& 0&$ (9.8 \pm 0.1)\%$&$ (5.8 \pm 0.2)\%$\\\hline
%&&&\multicolumn{2}{c}{UC double ratios}\\ 
%$a_U$ & $a_L$& $a_C$& $A_0$& $A_{12}$\\ \hline
%+5 &-5& -5 &$ (9.8 \pm 0.1)\%$&$ (6.0 \pm 0.2)\%$\\ 
%10 &0& 0&$ (9.8 \pm 0.1)\%$&$ (5.9 \pm 0.2)\%$\\ \hline
%\end{tabular}

\begin{tabular}{ccccccc}\hline
\multicolumn{3}{c}{weights  (\%)}&\multicolumn{2}{c}{UL double ratios (\%)}&\multicolumn{2}{c}{UC double ratios (\%)}\\
$a_U$ & $a_L$& $a_C$& $A_0$& $A_{12}$              & $A_0$& $A_{12}$                      \\ \hline
+5 &-5& -5 &$9.8 \pm 0.1$&$5.9 \pm 0.2$ &  $9.8 \pm 0.1$&$6.0 \pm 0.2$\\ 
10 &0& 0&$9.8 \pm 0.1$&$5.8 \pm 0.2$ &    $9.8 \pm 0.1$&$5.9 \pm 0.2$\\\hline
\end{tabular}

\end{table}

\subsection{Double ratios and subtraction of normalized yields}
The method of building double ratios, as described above, cancels possible acceptance effects as well as radiative effects to leading order. However, higher orders in the expansion of the radiative term might still remain and could affect the results if they were large. A second method exists that will cancel the radiative terms, but not necessarily the acceptance effects. If one subtracts the normalized yields for one charge combination from those of another charge combination,
\begin{eqnarray}
S^{UL}_\alpha = &:=&\frac{N^U_{\alpha}(\beta_\alpha)}{\langle N^U_{\alpha} \rangle} - \frac{N^C_{\alpha}(\beta_\alpha)}{\langle N^C_{\alpha} \rangle} \\
S^{UC}_\alpha = &:=&\frac{N^U_{\alpha}(\beta_\alpha)}{\langle N^U_{\alpha} \rangle} - \frac{N^C_{\alpha}(\beta_\alpha)}{\langle N^C_{\alpha} \rangle},
\end{eqnarray}
 one is sensitive only to the Collins asymmetry and possible acceptance effects. Again these yields are fitted by a constant and a cosine modulation $S^{ij}_\alpha = A^{ij}_\alpha \cos(\beta_\alpha) + B^{ij}_\alpha$, with $i,j \in \mathrm{U,L,C}$ where the constant fit parameter $B_\alpha^{ij}$ now is consistent with zero.
As a systematic study the results obtained from the double ratio method and the subtraction method have been compared and the absolute value of their differences has been assigned as a systematic error on the reconstruction method. As summarized in Table \ref{tab:systematics} these differences are small. The magnitude of the systematic uncertainty  in the $A^{UL}_{0}$ results ranges from less than 0.01\% in the bins with small fractional energies to 0.26\% in the highest $z$ bin where the smaller statistics might drive the differences. 
\subsection{$\pi^+\pi^+/\pi^-\pi^-$ double ratio tests\label{sec:chargedratios}}  
Another possible source of systematic errors could be a charge dependence of the detector response, which would manifest itself in azimuthal asymmetries. One can test this by probing double ratios of positively charged pion pairs over negatively charged pion pairs. Neither the Collins asymmetry nor the radiative effects depend on the pion charges. Therefore, one expects zero asymmetries. Again we fit the $A_\alpha$ results for all $z$ bins by a constant and assign the result together with its statistical error as a systematic uncertainty. The magnitude of the absolute systematic uncertainty is 0.05\% as summarized in Table \ref{tab:systematics}.
\subsection{$\tau^+\tau^-$ contribution to the asymmetries}
Background from $\tau$ pairs, which survive the selection, may give rise to fake asymmetries. To estimate this effect one has to use a data set that contains mostly $e^+e^-\rightarrow\tau^+\tau^-$ events. 
The $\tau$ content is enhanced by requiring a visible energy less than 10 GeV. In this sample about one third of the events originates from $\tau$ pairs while the tau contributions in the real data for the combined $z$ binning (symmetric $z_1,z_2$ were combined into one bin) are:
\begin{center}
\begin{tabular}{ccccccccccc}\hline
$z$ bin & 0 & 1 & 2 & 3 & 4 & 5 & 6 & 7 & 8 & 9 \\ \hline
$\#\pi_{\tau} /\# \pi_{all}$ &1\% &2\% &2\% &4\% &2\% &3\% &5\% &4\% &7\%&8\%  \\ \hline 
\end{tabular}
\end{center}
When analyzing the $\tau$ enhanced data in the same way as the full data sample one finds the following asymmetries: $A^{UL}=(0.179\pm0.300)\%$ for the $\cos2\phi_0$ method and  $A^{UL}=(0.750\pm0.300)\%$ for the $\cos(\phi_1 +\phi_2)$ method.
These values are consistent with zero within $2\sigma$ but consistently slightly positive; these results can be explained by the fairly large residual $e^+e^-\rightarrow q\overline{q}, q\in{udsc}$ contribution in the $\tau$ enhanced data sample. As a consequence it is not necessary to correct the measured asymmetries for this contribution. No systematic uncertainty has been assumed.
\subsection{Uncertainties due to particle identification (PID)}
While the applied pion selection only admits less than 10\% of misidentified pion pairs, the possible contributions from this misidentified background has to be checked. 
For this purpose a tighter pion likelihood selection has been applied, where the pion-kaon separation requirement was tightened to $\mathcal{L}(\pi)/[\mathcal{L}(K) + \mathcal{L}(\pi)] > 0.9 $. No significant changes have been observed. The small differences were added to the systematic uncertainties, which range from 0.01\% to 0.05\% (see Table \ref{tab:systematics}). 
\subsection{Uncertainties due to higher harmonics in the fit}
Furthermore one has to ensure the robustness of the fit. For this purpose one compares the cosine moments taken from the fit described above with the cosine moments when including an additional sine modulation and a cosine modulation of twice the argument as free parameters. No significant changes are observed and therefore no systematic uncertainty has been assumed.
In addition, varying the number of bins in azimuthal angle $\beta_\alpha$ between 6 and 24 gave stable results for the extracted $\cos\beta_\alpha$ moments. 
\subsection{Charm contributions and correction\label{sec:charm}}
\begin{table*}[th]
\begin{ruledtabular}
\begin{center}
\caption{\label{tab:dilution}Relative charm, {\it B} meson and uds contributions for the $z_1$,$z_2$ binning in the selected data sample and the $D^*$ sample in \%.}
\begin{tabular}{c c c c  c  c c  c}
& & \multicolumn{3}{c}{Data sample} & \multicolumn{3}{c}{$D^*$ sample}\\ \hline
$z_1$&$z_2$&$\#charm/\#all$&$\#B/\#all$&$\#uds/\#all$&$\#charm/\#all$&$\#B/\#all$&$\#uds/\#all$ \\ \hline
$[0.2,0.3]$&$[0.2,0.3]$& $ 24.88\pm  0.08 $&  $ 1.62\pm  0.02 $&  $ 73.50\pm  0.09 $&  $ 41.11\pm  0.12 $&  $ 2.11\pm  0.04 $&  $ 56.78\pm  0.12 $\\
$[0.2,0.3]$&$[0.3,0.5]$& $ 20.42\pm  0.09 $&  $ 1.66\pm  0.03 $&  $ 77.92\pm  0.10 $&  $ 39.21\pm  0.08 $&  $ 2.31\pm  0.03 $&  $ 58.48\pm  0.09 $\\
$[0.2,0.3]$&$[0.5,0.7]$& $ 13.72\pm  0.20 $&  $ 0.00\pm  0.00 $&  $ 86.28\pm  0.20 $&  $ 40.48\pm  0.20 $&  $ 0.01\pm  0.00 $&  $ 59.50\pm  0.20 $\\
$[0.2,0.3]$&$[0.7,1.0]$& $  2.88\pm  0.29 $&  $ 0.00\pm  0.00 $&  $ 97.12\pm  0.29 $&  $ 26.16\pm  0.91 $&  $ 0.00\pm  0.00 $&  $ 73.84\pm  0.91 $\\
$[0.3,0.5]$&$[0.2,0.3]$& $ 20.42\pm  0.09 $&  $ 1.72\pm  0.03 $&  $ 77.86\pm  0.10 $&  $ 39.19\pm  0.08 $&  $ 2.35\pm  0.03 $&  $ 58.46\pm  0.09 $\\
$[0.3,0.5]$&$[0.3,0.5]$& $ 16.59\pm  0.11 $&  $ 1.46\pm  0.03 $&  $ 81.94\pm  0.11 $&  $ 34.62\pm  0.08 $&  $ 2.10\pm  0.02 $&  $ 63.28\pm  0.08 $\\
$[0.3,0.5]$&$[0.5,0.7]$& $ 10.36\pm  0.21 $&  $ 0.00\pm  0.00 $&  $ 89.64\pm  0.21 $&  $ 31.06\pm  0.20 $&  $ 0.00\pm  0.00 $&  $ 68.93\pm  0.20 $\\
$[0.3,0.5]$&$[0.7,1.0]$& $  1.90\pm  0.27 $&  $ 0.00\pm  0.00 $&  $ 98.10\pm  0.27 $&  $ 13.32\pm  0.64 $&  $ 0.04\pm  0.04 $&  $ 86.64\pm  0.64 $\\
$[0.5,0.7]$&$[0.2,0.3]$& $ 13.20\pm  0.20 $&  $ 0.00\pm  0.00 $&  $ 86.80\pm  0.20 $&  $ 40.28\pm  0.20 $&  $ 0.00\pm  0.00 $&  $ 59.72\pm  0.20 $\\
$[0.5,0.7]$&$[0.3,0.5]$& $ 10.17\pm  0.21 $&  $ 0.00\pm  0.00 $&  $ 89.83\pm  0.21 $&  $ 31.49\pm  0.20 $&  $ 0.01\pm  0.00 $&  $ 68.50\pm  0.20 $\\
$[0.5,0.7]$&$[0.5,0.7]$& $  5.69\pm  0.38 $&  $ 0.00\pm  0.00 $&  $ 94.31\pm  0.38 $&  $ 24.52\pm  0.46 $&  $ 0.00\pm  0.00 $&  $ 75.48\pm  0.46 $\\
$[0.5,0.7]$&$[0.7,1.0]$& $  1.60\pm  0.56 $&  $ 0.00\pm  0.00 $&  $ 98.40\pm  0.56 $&  $ 10.99\pm  1.32 $&  $ 0.00\pm  0.00 $&  $ 89.01\pm  1.32 $\\
$[0.7,1.0]$&$[0.2,0.3]$& $  2.81\pm  0.28 $&  $ 0.00\pm  0.00 $&  $ 97.19\pm  0.28 $&  $ 25.09\pm  0.92 $&  $ 0.09\pm  0.06 $&  $ 74.82\pm  0.93 $\\
$[0.7,1.0]$&$[0.3,0.5]$& $  2.33\pm  0.30 $&  $ 0.00\pm  0.00 $&  $ 97.67\pm  0.30 $&  $ 14.00\pm  0.66 $&  $ 0.00\pm  0.00 $&  $ 86.00\pm  0.66 $\\
$[0.7,1.0]$&$[0.5,0.7]$& $  1.51\pm  0.57 $&  $ 0.00\pm  0.00 $&  $ 98.49\pm  0.57 $&  $  9.60\pm  1.20 $&  $ 0.00\pm  0.00 $&  $ 90.40\pm  1.20 $\\
$[0.7,1.0]$&$[0.7,1.0]$& $  0.00\pm  0.00 $&  $ 0.00\pm  0.00 $&  $100.00\pm  0.00 $&  $ 10.00\pm  4.74 $&  $ 0.00\pm  0.00 $&  $ 90.00\pm  4.74 $\\
\end{tabular}
\end{center}
\end{ruledtabular}
\end{table*}
The contribution from $e^+e^-\rightarrow c\overline{c}$ amounts to about 40\% of the total quark-antiquark production cross-section. Due to weak decays of charmed hadrons, which can introduce azimuthal asymmetries not originating from the Collins effect, the results have to be corrected for this contribution.
\subsubsection{Charm enhanced data sample}
The decays of charmed hadrons are well described in the MC and, one can derive the relative contributions $D:= \frac{N_{\rm charm}}{N_{\rm all}}|_{\rm data}$ of pion pairs in the data sample. 
Additionally one can select a charm-enhanced data sample, where one requires a $D\pi$ pair invariant mass ($m(D\pi)$) consistent with the mass of the $D^*$ meson, corresponding to decays $D^{*+}\rightarrow D^0\pi^+$ (and charge-conjugated process). These decays are selected with a high purity by requiring (in addition to the selection criteria already described) the invariant mass difference $\Delta m= m(D\pi)- m(D)$ consistent with the nominal mass difference of $D^*$ and $D$ mesons \cite{PDG}. One can calculate the charm quark-antiquark pair contributions to this data sample as $d:= \frac{N_{\rm charm}}{N_{\rm all}}\big|_{D^*}$ (see Table \ref{tab:dilution}). 
Measuring the double ratio asymmetries $A_{\alpha}^{ij}$ simultaneously in the default as well as in the $D^*$ meson enhanced data samples and assuming that the asymmetries from the charm events are the same in both samples, one obtains two equations for the two separate data samples:
\begin{eqnarray}
\label{eq:auds}A_{\rm uds}&=&\frac{1}{1-D} A_{\rm all} - \frac{D}{1-D}A_{\rm charm} \\
\label{eq:acharm}A_{\rm charm}&=&\frac{1}{d} A_{D^*} - \frac{1-d}{d}A_{\rm uds} \quad ,
\end{eqnarray}
where $A_{\rm all}$ are the measured asymmetries in the default data sample, $A_{D^*}$ those of the charm enhanced data sample and $A_{\rm uds}$ and $A_{\rm charm}$ the true asymmetries of uds or charm quarks, respectively. 
As a result one obtains corrected asymmetries for the light quarks. The statistical errors of both samples were propagated to obtain the statistical errors for $A_{\rm uds}$ (and $A_{\rm charm}$). This leads to comparatively large statistical errors for the charm corrected asymmetries due to the low statistics in the charm enhanced data sample.
\subsection{Bottom background}
Due to the thrust selection and the requirement on the fractional energy of the tracks only a small fraction of pion pairs from  $B\overline{B}$ decays appears in the data sample. Only the bins with fractional energies below 0.5 can have a B-contribution. As a consequence the contributions $B:=\frac{N_{B\overline{B}}}{N_{\rm all}}$ is at most 2.5\% as can be seen from Table \ref{tab:dilution}. Correspondingly the systematic uncertainties due to {\it B} meson decays in data are negligible compared to other uncertainties. 
In correcting the light quark asymmetries for the charm contribution the small contributions from bottom background were taken into account by replacing $1-D$ by $1- D -B$ and similarly $1-d$ by $1-d-b$. The bottom quark asymmetry has been assumed to be zero. 
\subsection{Beam polarization measurements\label{sec:bpol}}
Another important study is a test of the polarizations of the electron and positron beams. The natural beam polarization in electron storage rings is a transverse polarization with the electron spins aligned with the magnetic bending fields. This is a consequence of the emission of synchrotron radiation as detailed in a paper by Sokolov and Ternov \cite{sokolov}.
In the KEKB storage ring the polarization will be destroyed by the large tune shifts in the beam-beam interactions. A measurement of any residual polarization is, however, in order.
\begin{figure}[th]
\begin{center}
\includegraphics[width=6cm]{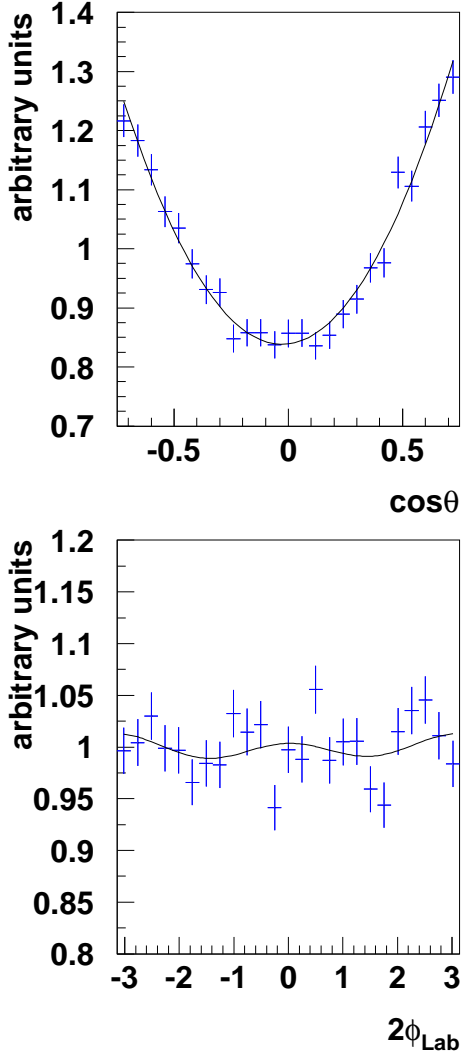}
\caption{Angular distributions of $\cos\theta_{\rm CMS}$ (top plot) and $2\phi_{LAB}$ (bottom plot) for muon pairs.\label{fig:mumu_fit}}
\end{center}
\end{figure}
While the process $e^+e^-\rightarrow \gamma \gamma$ appears to be the most sensitive to a transverse beam polarization, the process $e^+e^-\rightarrow \mu^+\mu^-$ is experimentally easier to asses. For muon pair events the $\cos(\theta)$ and $\phi_{LAB}$ distributions are analyzed, where $\theta$ is the polar angle between the electron and the muon axis in the CMS and $\phi_{LAB}$ is the azimuthal angle of the muon in the laboratory frame around the z-axis. Both distributions are normalized to the average bin content. In the case of the $\cos(\theta)$ distribution this procedure was performed only in the range $[-0.75,0.75]$ since at larger angles acceptance effects dominate.
The $\cos(\theta)$ distribution is then fitted with the functional form $a(1+b\cos^2\theta+c\cos\theta)$. The unpolarized cross section should contain a $\cos^2(\theta)$ term equal to unity and a $\cos(\theta)$ moment of a few percent, due to photon-Z interference and higher order QED terms. An example of these fits is shown in Fig.~\ref{fig:mumu_fit}. The $\theta$ moments show the expected modulations of $b=1.02\pm0.04$ and $c=0.056\pm0.0015$ and show that the muon selection and detector performance is reliable for muon pairs.
The $\phi_{LAB}$ distribution is fitted by the function $a+b\cos(\phi_{LAB})+ c\sin(\phi_{LAB}) + d \cos(2\phi_{LAB})$. An example of such a fit is shown in Fig.~\ref{fig:mumu_fit}. Any nonzero $\phi_{LAB}$ moment would be a sign of a nonzero beam polarization. The $\cos 2\phi_{LAB}$ moment is especially sensitive since it is proportional to the product of the two beam polarizations \cite{oldspear}.
The fits to the $2\phi_{LAB}$ distributions were performed either by taking all the muon pairs of one run or by dividing individual runs into several time bins to study the time behavior of a possible beam polarization. Run periods before and after the implementation of continuous injection as a mode of KEKB operation \cite{kekb} were also compared. The $\phi_{LAB}$ moments are consistent with zero in all time periods. No build-up of polarization could be observed. Therefore one can conclude that no significant beam polarization exists in this data sample and no systematic error is assigned.
\subsection{Correlation studies}
The statistical errors obtained from the fits to the double ratios may be influenced by correlations among individual angular and ($z_1$, $z_2$) bins. The correlations can arise since different pairs composed of the same pion are allowed to contribute to the distributions. The effect on the statistical errors was tested by dividing the large reweighted MC sample into a large number of subsamples. The same fit as for the data was performed using each of the MC subsamples. The width of the distribution of the fit results is found to be in agreement with the statistical error of the fit performed on the full MC sample.
\subsection{Internal consistency test\label{sec:consistency}}
\begin{figure}[th]
\begin{center}
\includegraphics[width=7cm]{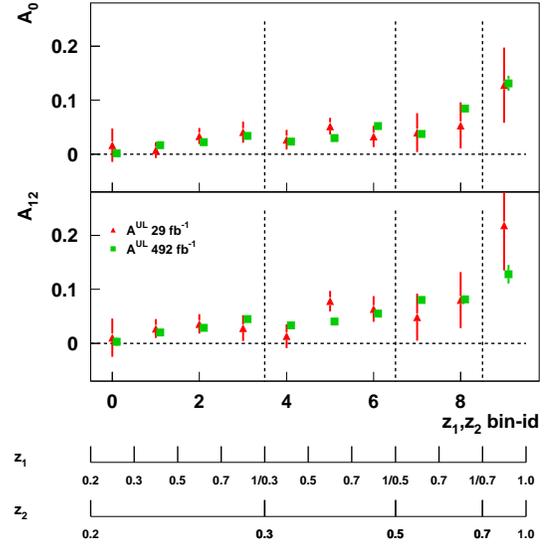}
\caption{Comparison of the $A^{UL}$ asymmetries for the published off-resonance data sample (triangles) and the on-resonance data sample (squares). The upper plot shows the $A_0$ asymmetries, the lower plot the $A_{12}$ asymmetries as a function of $z$.\label{fig:contres}}
\end{center}
\end{figure}

Before combining the published data \cite{belleprl} obtained from the off-resonance sample and the additional on-resonance data, the consistency of the two data sets has to be tested. The charm corrected results from the published $29.1$ fb$^{-1}$ data sample and the results from the $492$ fb$^{-1}$ data are displayed in Fig.~\ref{fig:contres}; the results are in good agreement in all combined z-bins. For this comparison the correction factors for the published data, which are based on the generated thrust axis instead of the quark axis were applied. The overall $\chi^2 = \sum_{i=1,10} \frac{(A_i(set 1) - A_i(set 2))^2}{\Delta A_i^2(set 1) + \Delta A_i^2(set 2)}$ of the two data samples per degree of freedom is 0.54 for the $A_0^{UL}$ results and 0.83 for the $A_{12}^{UL}$. Therefore it is possible to combine the off-resonance and on-resonance data sets. The differences in the CMS energy are automatically accounted for by considering fractional energies $z$.

\begin{figure}[th]
\begin{center}
\includegraphics[width=7cm]{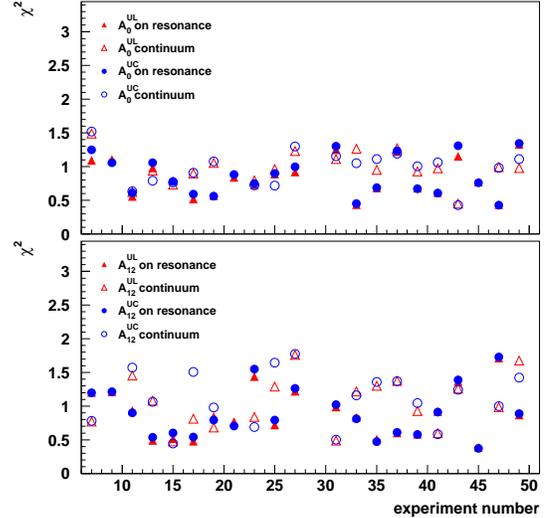}
\caption{$\chi^2$ distribution per degree of freedom of the $A^{UL}$ (triangles) and $A^{UC}$ (circles) asymmetries as functions of the experiment number for the $A_0$ (top) and the $A_{12}$ (bottom) asymmetries. The open symbols represent the off-resonance data samples, the full symbols the on-resonance data samples.\label{fig:chivsexp}}
\end{center}
\end{figure}

\begin{figure}[th]
\begin{center}
\includegraphics[width=7cm]{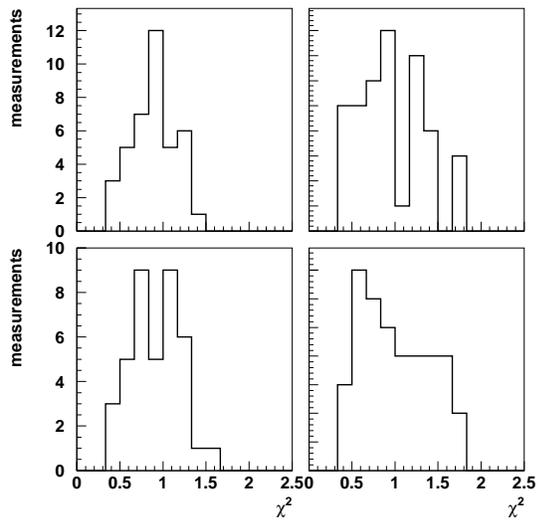}
\caption{$\chi^2$ distribution  per degree of freedom of the $A^{UL}$ (top plots) and $A^{UC}$ (bottom plots) asymmetries for the $A_0$ (left plots) and the $A_{12}$ method (right plots).\label{fig:chi2}}
\end{center}
\end{figure}

An additional test to compare the results of different data taking periods for on-resonance data and off-resonance data and verify that these are consistent as well as independent of time. At Belle the data is naturally divided into periods of several-months' data taking, called experiments with odd, increasing numbers starting at 7. As a reference the double ratio results for the complete data set (experiments 7-49) for both types of double ratios ( UL and UC) and extraction methods (0 and 12) have been taken. For each experiment number the $\chi^2$ value per degree of freedom relative to the combined result is calculated. The $\chi^2$ values are presented in Fig.~\ref{fig:chivsexp} as a function of the experiment number. No systematic trend for any of the data samples or methods can be observed. The distributions of $\chi^2$ per degree of freedom are also displayed in Fig.~\ref{fig:chi2}, where one sees that they scatter around the central value of 1.\\

\begin{figure}[th]
\begin{center}
\includegraphics[width=8cm]{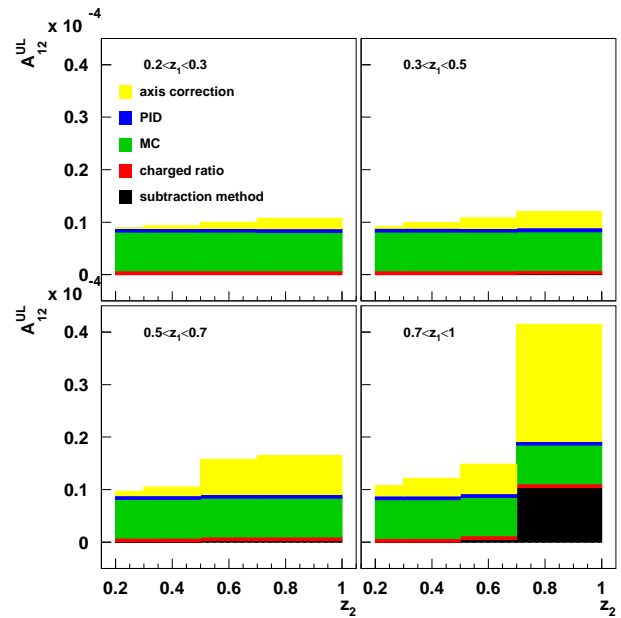}
\caption{\label{fig:systematics}Squares of the contributions to the systematic error for the $A_{12}^{UL}$ asymmetries as a function of $z_1$ and $z_2$. The lowermost bars correspond to the contribution from the differences between the double ratio and subtraction reconstruction methods. The next bars correspond to the charged pion pair ratios, the next to the MC double ratios, then the contribution due to particle identification and the uppermost bar corresponds to the uncertainties in correcting the measured asymmetries for the quark-antiquark/thrust axis deviation.}
\end{center}
\end{figure}

In summary a number of possible sources of uncertainties in the asymmetry extraction have been studied and their contributions have been evaluated. As can be seen in Fig.~\ref{fig:systematics} and Table \ref{tab:systematics} the errors are dominated by the detector effects on the double ratios and the statistical uncertainties on them. In general, most of the systematic uncertainties have significantly decreased in comparison to the previously published data \cite{belleprl} as the statistics of the data that are used to evaluate some of the systematic uncertainties also increased by a factor of almost 20.

\begin{table*}[hb]
\begin{ruledtabular}
\begin{center}
\caption{\label{tab:systematics}Absolute contributions to the systematic error on the $A_{\alpha}^{UL}$ and $A_{\alpha}^{UC}$ results for the $z_1,z_2$ binning. The different columns contain the systematic errors arising from subtraction method (1), ratios of positively over negatively charged pion pairs (2), double ratios from MC (3), uncertainties due to particle identification (4) and due to underestimation in reweighted MC (5) as described in the text.}
\begin{tabular}{ccc c  c  c cc cc cc}
 &&\multicolumn{5}{c}{$A^{UL}_0$}&\multicolumn{5}{c}{$A^{UL}_{12}$} \\ 
$z_1$&$z_2$&1 & 2 & 3 & 4 & 5 & 1 & 2 & 3 & 4 & 5 \\ \hline \hline
$[0.2,0.3]$&$[0.2,0.3]$&  0.0001 &  0.0005&  0.0022&  0.0002&0.0002&  0.0001&   0.0005&  0.0016&  0.0005&0.0002\\
$[0.2,0.3]$&$[0.3,0.5]$&  0.0001 &  0.0005&  0.0022&  0.0002&0.0009&  0.0001&   0.0005&  0.0016&  0.0005&0.0007\\
$[0.2,0.3]$&$[0.5,0.7]$&  0.0001 &  0.0005&  0.0022&  0.0002&0.0012&  0.0001&   0.0005&  0.0016&  0.0005&0.0011\\
$[0.2,0.3]$&$[0.7,1.0]$&  0.0000 &  0.0005&  0.0022&  0.0002&0.0019&  0.0000&   0.0005&  0.0016&  0.0005&0.0014\\
$[0.3,0.5]$&$[0.2,0.3]$&  0.0001 &  0.0005&  0.0022&  0.0002&0.0008&  0.0001&   0.0005&  0.0016&  0.0005&0.0006\\
$[0.3,0.5]$&$[0.3,0.5]$&  0.0001 &  0.0005&  0.0022&  0.0002&0.0012&  0.0001&   0.0005&  0.0016&  0.0005&0.0011\\
$[0.3,0.5]$&$[0.5,0.7]$&  0.0000 &  0.0005&  0.0022&  0.0002&0.0015&  0.0001&   0.0005&  0.0016&  0.0005&0.0014\\
$[0.3,0.5]$&$[0.7,1.0]$&  0.0004 &  0.0005&  0.0022&  0.0002&0.0028&  0.0002&   0.0005&  0.0016&  0.0005&0.0018\\
$[0.5,0.7]$&$[0.2,0.3]$&  0.0001 &  0.0005&  0.0022&  0.0002&0.0012&  0.0001&   0.0005&  0.0016&  0.0005&0.0009\\
$[0.5,0.7]$&$[0.3,0.5]$&  0.0000 &  0.0005&  0.0022&  0.0002&0.0016&  0.0001&   0.0005&  0.0016&  0.0005&0.0013\\
$[0.5,0.7]$&$[0.5,0.7]$&  0.0003 &  0.0005&  0.0022&  0.0002&0.0019&  0.0003&   0.0005&  0.0016&  0.0005&0.0026\\
$[0.5,0.7]$&$[0.7,1.0]$&  0.0004 &  0.0005&  0.0022&  0.0002&0.0048&  0.0003&   0.0005&  0.0016&  0.0005&0.0027\\
$[0.7,1.0]$&$[0.2,0.3]$&  0.0002 &  0.0005&  0.0022&  0.0002&0.0015&  0.0000&   0.0005&  0.0016&  0.0005&0.0014\\
$[0.7,1.0]$&$[0.3,0.5]$&  0.0001 &  0.0005&  0.0022&  0.0002&0.0024&  0.0000&   0.0005&  0.0016&  0.0005&0.0018\\
$[0.7,1.0]$&$[0.5,0.7]$&  0.0001 &  0.0005&  0.0022&  0.0002&0.0035&  0.0004&   0.0005&  0.0016&  0.0005&0.0024\\
$[0.7,1.0]$&$[0.7,1.0]$&  0.0026 &  0.0005&  0.0022&  0.0002&0.0069&  0.0019&   0.0005&  0.0016&  0.0005&0.0047\\\hline\hline \\
&&\multicolumn{5}{c}{$A^{UC}_0$}&\multicolumn{5}{c}{$A^{UC}_{12}$} \\
$z_1$&$z_2$&1 & 2 & 3 & 4 & 5 & 1 & 2 & 3 & 4 & 5 \\ \hline \hline 
$[0.2,0.3]$&$[0.2,0.3]$&  0.0000 &  0.0005&  0.0010&  0.0001&0.0001&  0.0000&   0.0005&  0.0007&  0.0002&0.0001\\
$[0.2,0.3]$&$[0.3,0.5]$&  0.0000 &  0.0005&  0.0010&  0.0001&0.0004&  0.0000&   0.0005&  0.0007&  0.0002&0.0003\\
$[0.2,0.3]$&$[0.5,0.7]$&  0.0000 &  0.0005&  0.0010&  0.0001&0.0005&  0.0000&   0.0005&  0.0007&  0.0002&0.0005\\
$[0.2,0.3]$&$[0.7,1.0]$&  0.0000 &  0.0005&  0.0010&  0.0001&0.0008&  0.0000&   0.0005&  0.0007&  0.0002&0.0006\\
$[0.3,0.5]$&$[0.2,0.3]$&  0.0000 &  0.0005&  0.0010&  0.0001&0.0003&  0.0000&   0.0005&  0.0007&  0.0002&0.0002\\
$[0.3,0.5]$&$[0.3,0.5]$&  0.0000 &  0.0005&  0.0010&  0.0001&0.0005&  0.0000&   0.0005&  0.0007&  0.0002&0.0005\\
$[0.3,0.5]$&$[0.5,0.7]$&  0.0000 &  0.0005&  0.0010&  0.0001&0.0006&  0.0000&   0.0005&  0.0007&  0.0002&0.0006\\
$[0.3,0.5]$&$[0.7,1.0]$&  0.0001 &  0.0005&  0.0010&  0.0001&0.0011&  0.0000&   0.0005&  0.0007&  0.0002&0.0006\\
$[0.5,0.7]$&$[0.2,0.3]$&  0.0000 &  0.0005&  0.0010&  0.0001&0.0005&  0.0000&   0.0005&  0.0007&  0.0002&0.0004\\
$[0.5,0.7]$&$[0.3,0.5]$&  0.0000 &  0.0005&  0.0010&  0.0001&0.0007&  0.0000&   0.0005&  0.0007&  0.0002&0.0005\\
$[0.5,0.7]$&$[0.5,0.7]$&  0.0000 &  0.0005&  0.0010&  0.0001&0.0007&  0.0001&   0.0005&  0.0007&  0.0002&0.0010\\
$[0.5,0.7]$&$[0.7,1.0]$&  0.0001 &  0.0005&  0.0010&  0.0001&0.0016&  0.0001&   0.0005&  0.0007&  0.0002&0.0009\\
$[0.7,1.0]$&$[0.2,0.3]$&  0.0001 &  0.0005&  0.0010&  0.0001&0.0006&  0.0000&   0.0005&  0.0007&  0.0002&0.0006\\
$[0.7,1.0]$&$[0.3,0.5]$&  0.0000 &  0.0005&  0.0010&  0.0001&0.0009&  0.0001&   0.0005&  0.0007&  0.0002&0.0007\\
$[0.7,1.0]$&$[0.5,0.7]$&  0.0000 &  0.0005&  0.0010&  0.0001&0.0011&  0.0000&   0.0005&  0.0007&  0.0002&0.0007\\
$[0.7,1.0]$&$[0.7,1.0]$&  0.0004 &  0.0005&  0.0010&  0.0001&0.0015&  0.0000&   0.0005&  0.0007&  0.0002&0.0009\\
\end{tabular}
\end{center}
\end{ruledtabular}
\end{table*}

\section{Results}
The final results combine the 55 fb$^{-1}$ data sample taken at an energy of $10.52$ GeV and 492 fb $^{-1}$ of data taken on the $\Upsilon(4S)$ resonance at $10.58$ GeV. Since the fractional energies $z_{1,2}$ are already normalized by the corresponding CM energies the two data sets have been combined. The double ratios have been evaluated and a fit was performed as described above. The asymmetries have been corrected for the charm contribution and were rescaled by the factors obtained by the weighted MC ($1.66\pm 0.04$ for the $A_{12}$ asymmetries, $ 1.11\pm0.05$ for the $A_0$ asymmetries). 
\subsection{Double ratio results}
\subsubsection{Double ratios versus fractional energies $z_1z_2$}
\begin{figure}[t]
\begin{center}
\includegraphics[width=8cm]{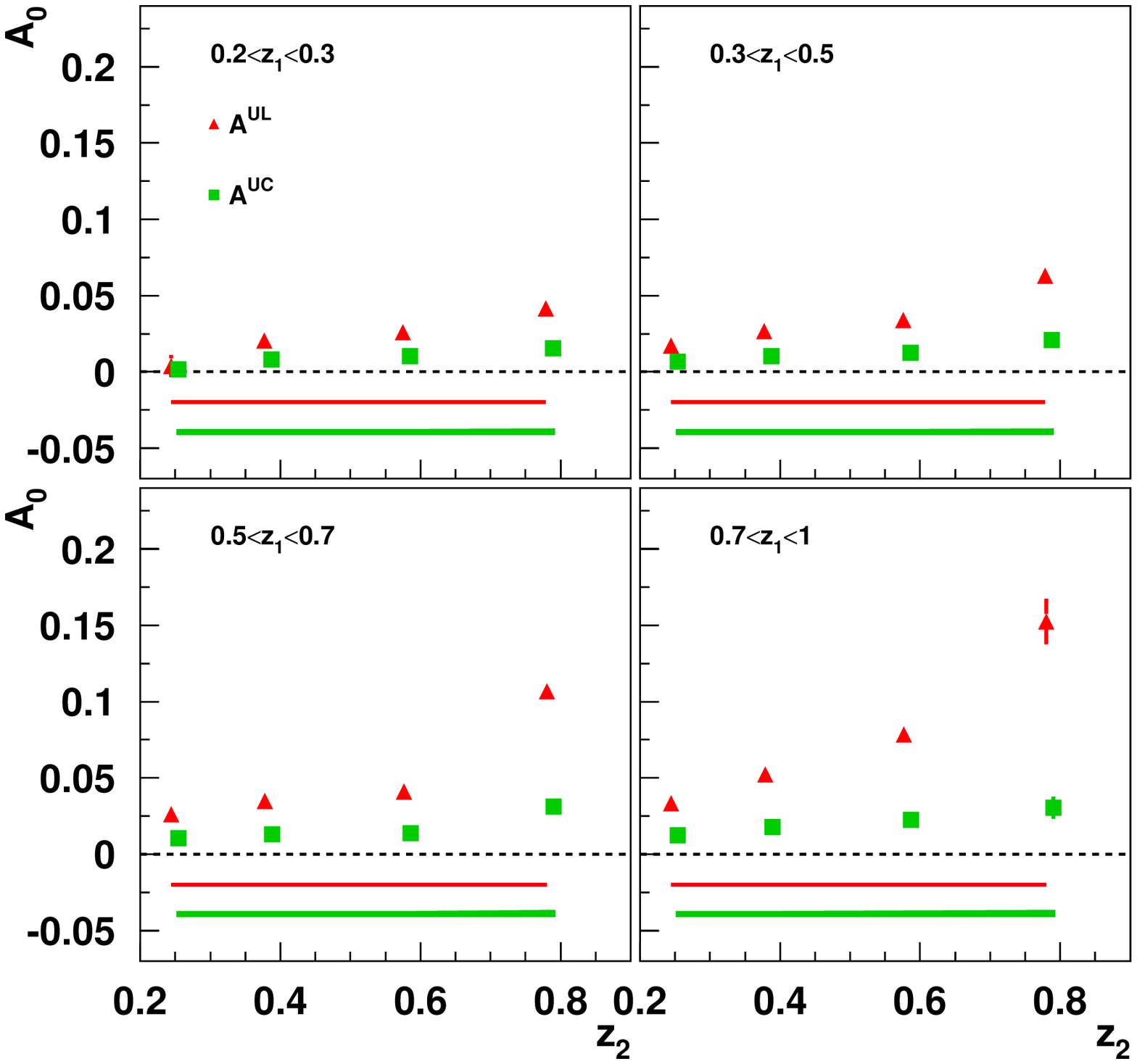}
\caption{\label{fig:finalpanela0}Light quark ({\it uds}) $A_{0}$ asymmetry parameters as a function of $z_2$ for 4 $z_1$ bins. The UL data are represented by triangles and the systematic error by the upper error band. The UC data are described by the squares and their systematic uncertainty by the lower error band.}
\end{center}
\end{figure}

\begin{figure}[t]
\begin{center}
\includegraphics[width=8cm]{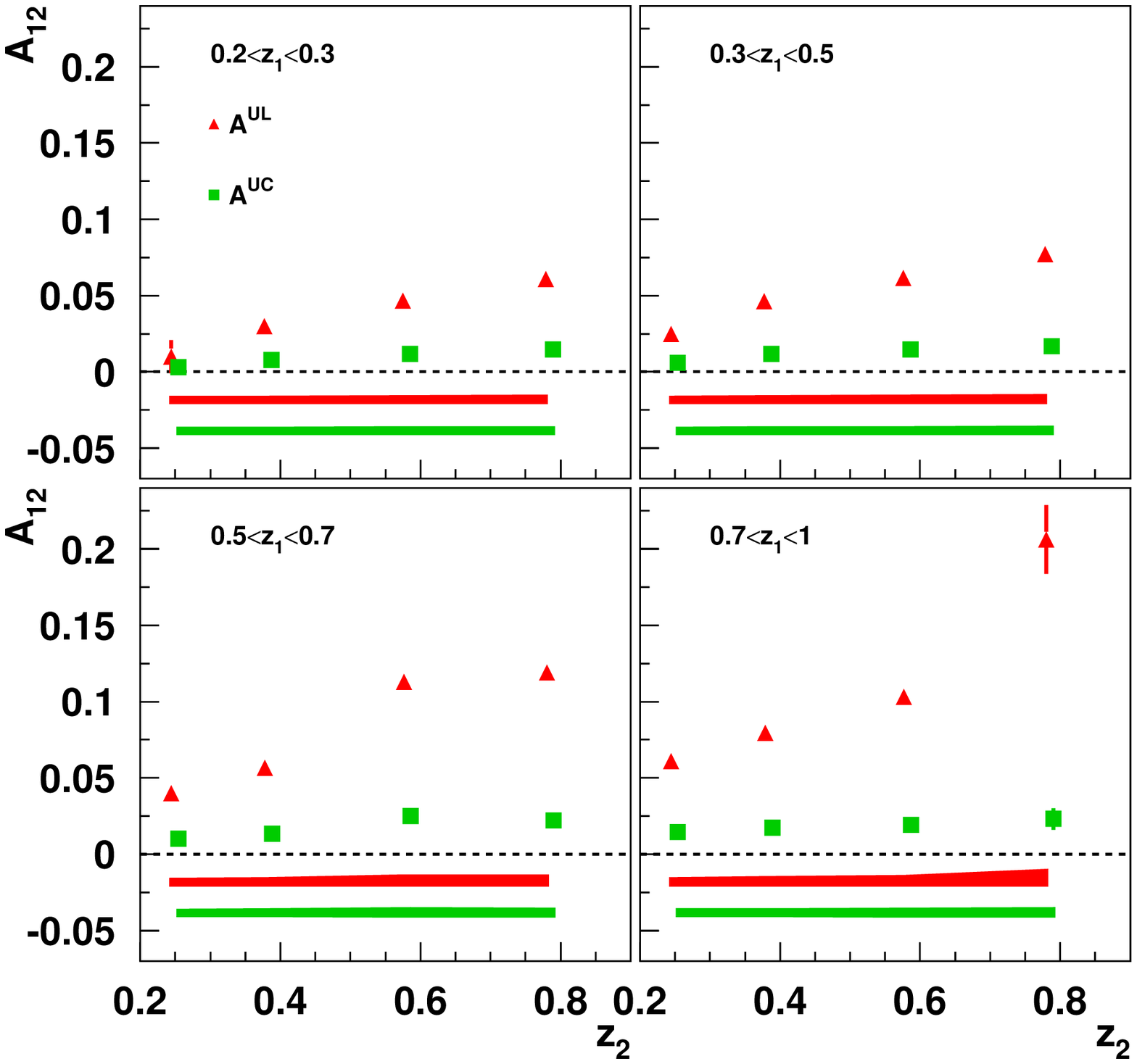}
\caption{\label{fig:finalpanela12}Light quark ({\it uds}) $A_{12}$ asymmetry parameters as a function of $z_2$ for 4 $z_1$ bins. The UL data are represented by triangles and the systematic error by the upper error band. The UC data are described by the squares and their systematic uncertainty by the lower error band.}
\end{center}
\end{figure}

The main results are the asymmetry parameters $A_0$ and $A_{12}$ for both types of double ratios (UL and UC) as a function of the fractional energies of the two hadrons. Figures \ref{fig:finalpanela0} and \ref{fig:finalpanela12} show these asymmetries where all $z_2$ bins for a given $z_1$ are displayed. The numerical values are give in Tables \ref{tab:dr} and \ref{tab:dr2}. One can clearly see the rising asymmetry in each plot as a function of $z_2$. The UC asymmetries are significantly smaller than the UL asymmetries but non-zero, which, given the different contributions of favored and disfavored fragmentation functions already suggests a large disfavored Collins fragmentation function with opposite sign to the favored one. This suggestion will be quantified in the next section. 

\subsubsection{Double ratios versus polar angle $\sin^2\theta/(1+\cos^2\theta)$}
\begin{figure}[t]
\begin{center}
\includegraphics[width=8cm]{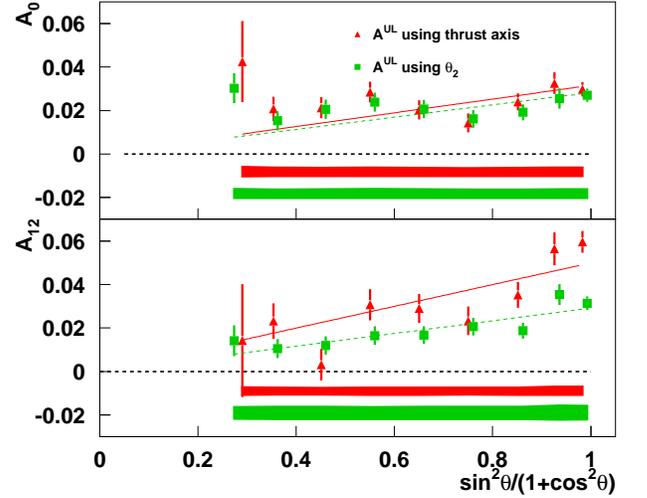}
\caption{\label{fig:finalsinth}Light quark (uds) $A^{UL}_{0}$ (top) and $A^{UL}_{12}$ (bottom) asymmetry parameters as a function of $\sin^2\theta/(1+\cos^2\theta)$, for $\theta_2$ (squares) and for $\hat{\mathbf{n}}_z$ (triangles). Linear fits are also displayed as dashed and continuous lines, respectively. The systematic error for $\theta_2$ case is represented by the lower, that for $\hat{\mathbf{n}}_z$ by the upper error band.}
\end{center}
\end{figure}

\begin{figure}[t]
\begin{center}
\includegraphics[width=8cm]{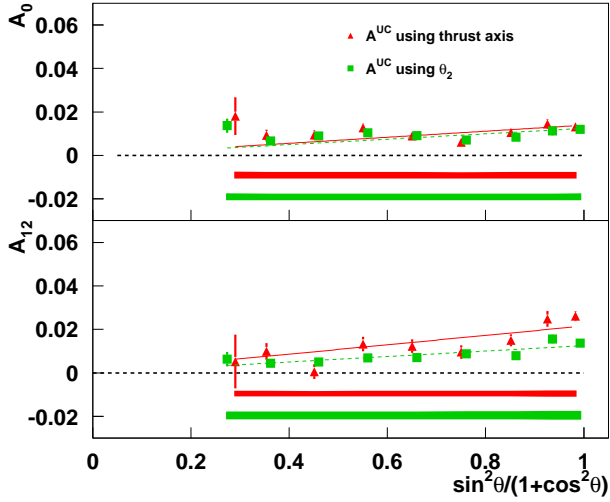}
\caption{\label{fig:finalsinth2}Light quark (uds) $A^{UC}_{0}$ (top) and $A^{UC}_{12}$ (bottom) asymmetry parameters as a function of $\sin^2\theta/(1+\cos^2\theta)$, for $\theta_2$ (squares) and for $\hat{\mathbf{n}}_z$ (triangles). Linear fit are also displayed as dashed and continuous lines, respectively. The systematic error for $\theta_2$ case is represented by the lower, that for $\hat{\mathbf{n}}_z$ by the upper error band.}
\end{center}
\end{figure}

Another way of presenting the results of the asymmetry measurements is based on the requirement for primordial transverse quark polarization. At leading order and ignoring $\gamma-Z$ interference transverse quark spins leads to a $\frac{\sin^2(\theta)}{1+\cos^2(\theta)}$ dependence of the asymmetries (see Eqs. \ref{eq:dr}, \ref{eq:dr2}), where $\theta$ can be represented by either the polar angle between the $e^+e^-$ and the thrust axis $\hat{\mathbf{n}}_z$, or by the polar angle $\theta_2$ of the 2$^{nd}$ hadron relative to the $e^+e^-$ axis. Figure \ref{fig:finalsinth} displays the $A_0$ and $A_{12}$ results for the UL double ratios as a function of $\frac{\sin^2{\theta}}{1+\cos^2(\theta)}$ while Fig.~\ref{fig:finalsinth2} displays the results for the UC double ratios. Both polar angles have been considered and each of them has been fit by a first order polynomial where the constant term has been set to zero. In both cases the results are consistent with a linear behavior. The results obtained with the thrust axis defining the polar angle can be described by the linear term only as the $\chi^2$ per degree of freedom of the fit changes only slightly when allowing the constant term to float, for example for the $A^{UL}_0$ result from 2.4 to 1.67 and from 2.56 to 2.35 for the $A_{12}^{UL}$ result. The $A_0$ results obtained with $\theta_2$ as the polar angle favor a nonzero constant term; when a constant term is included the reduced $\chi^2$ of the fit decreases significantly from 2.81 to 1.26 for the $A^{UL}_0$ result and from 2.57 to 1.22 for the $A_{0}^{UC}$ result. This can be explained by the fact that the thrust axis describes the original quark direction better than the 2$^{nd}$ hadron's polar angle, which receives some additional transverse momentum relative to the quark axis. 

\subsubsection{Double ratios versus $Q_T$ for high and low thrust data samples}
\begin{figure}[t]
\begin{center}
\includegraphics[width=8cm]{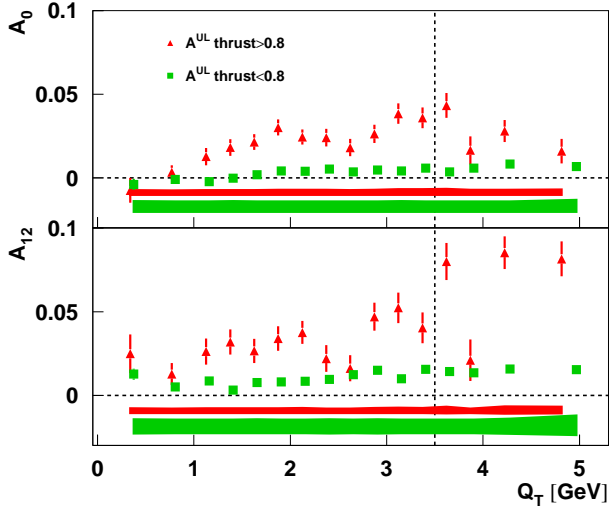}
\caption{\label{fig:finalqt}Light quark (uds) $A^{UL}_{0}$ (top) and $A^{UL}_{12}$ (bottom) asymmetry parameters as a function of $Q_T$, for events with $T>0.8$ (triangles) and asymmetries for events with $T<0.8$ not corrected for heavy quark contributions (squares). The vertical line represents the main data selection $Q_T<3.5$ GeV.}
\end{center}
\end{figure}

\begin{figure}[t]
\begin{center}
\includegraphics[width=8cm]{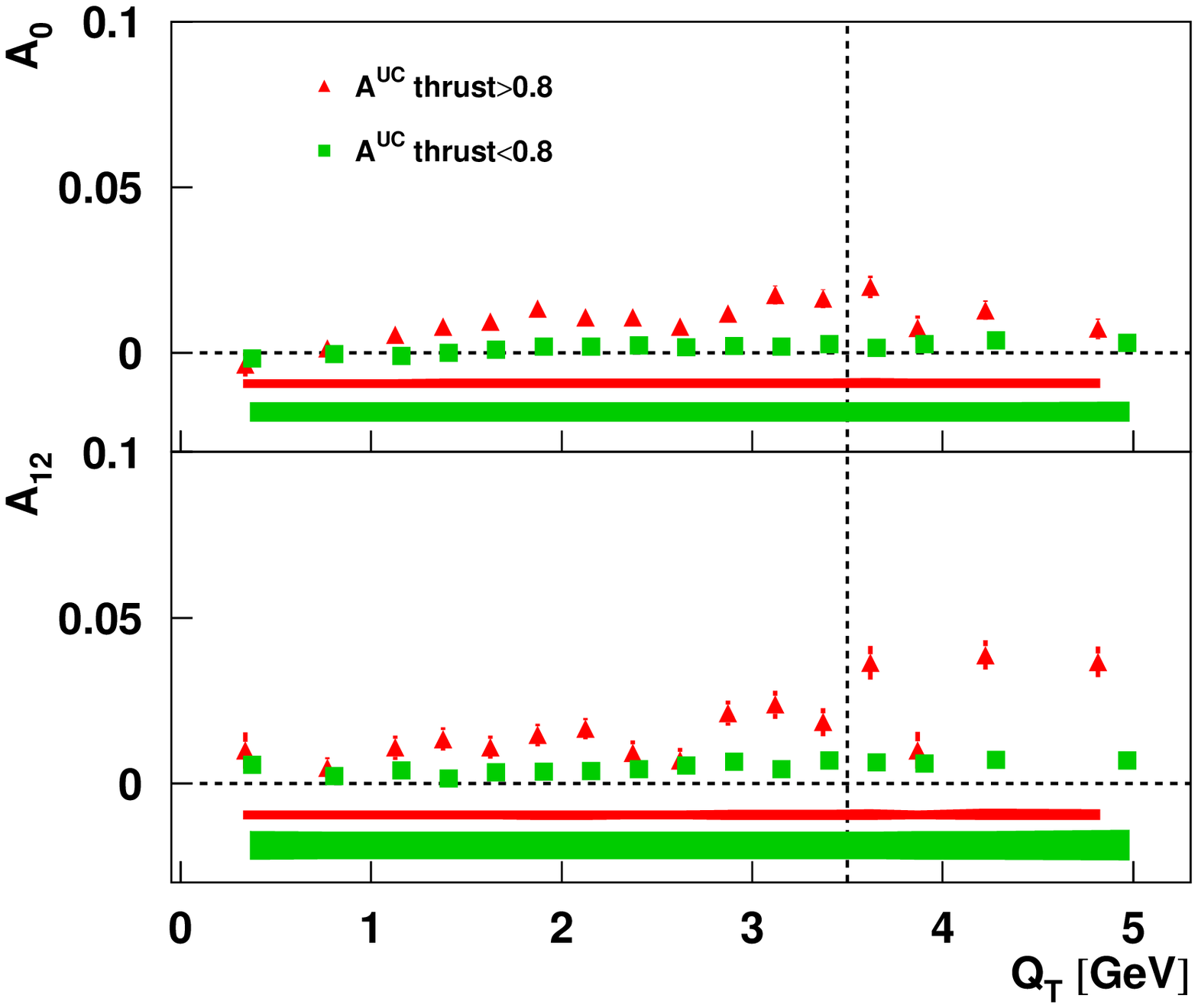}
\caption{\label{fig:finalqt2}Light quark (uds) $A^{UC}_{0}$ (top) and $A^{UC}_{12}$ (bottom) asymmetry parameters as a function of $Q_T$, for events with $T>0.8$ (triangles) and asymmetries for events with $T<0.8$ not corrected for heavy quark contributions (squares). The vertical line represents the main data selection $Q_T<3.5$ GeV.}
\end{center}
\end{figure}

The dependence of the asymmetries on the virtual photon momentum in the two-hadron center-of-mass frame is also of interest. The results are shown in Figs.~\ref{fig:finalqt} and \ref{fig:finalqt2}. In addition to the charm-corrected asymmetries the asymmetries for the reverse thrust selection $T<0.8 $ are displayed. The contributions of both charm quarks and by $\Upsilon(4S)$ decays are quite substantial in the reverse thrust selection sample and can add up to almost 70\% in the highest $Q_T$ bin. 
The results of the reverse thrust selection are displayed uncorrected for the charm and the $\Upsilon(4S)$ contributions. When comparing the reverse thrust selection for on and off-resonance data one sees that the $\Upsilon(4S)$ does give an additional contribution to the $A_{12}$ asymmetries. Nevertheless it is clearly visible that the asymmetries are significantly lower than in the main data selection. This is the expected behavior, since the asymmetries due to the Collins effect are smeared out for events with no clear two-jet structure.

\subsubsection{Charm asymmetries}
\begin{figure}[t]
\begin{center}
\includegraphics[width=7.5cm]{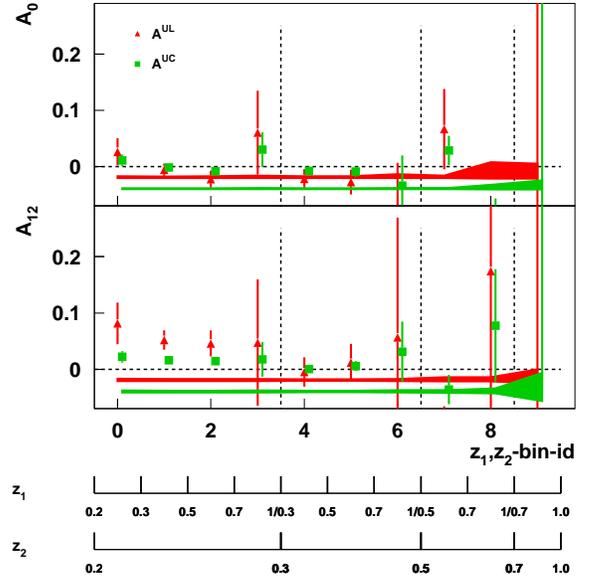}
\caption{\label{fig:charmfinal}Charm  asymmetry parameters $A_{0}$ and $A_{12}$ as a function of the combined z. The UL data are described by triangles, their systematic error being the top error band while the UC data are described by the squares and their systematics by the lower error band.}
\end{center}
\end{figure}

Eqns.~(\ref{eq:auds}) and (\ref{eq:acharm}) can also be solved for the asymmetries from the process $e^+e^-\rightarrow c\overline{c}$. The results as a function of the fractional energies $z_1$ and $z_2$ are displayed in Fig.~\ref{fig:charmfinal}. While a small asymmetry is visible at lower fractional energies it seems to be consistent with zero at larger fractional energies, although the statistical errors become large. The integrated results for the $A_0$ asymmetries are compatible with zero ($\langle A_{0}^{UL}\rangle = -0.011\pm0.007$ and $\langle A_{0}^{UC}\rangle = -0.004\pm0.003$), while the $A_{12}$ asymmetries are found to be slightly positive ($\langle A_{12}^{UL}\rangle = 0.037\pm0.011$ and $\langle A_{0}^{UC}\rangle = 0.012\pm0.003$).

\section{Parameterization of the Collins function\label{sec:param}}
After obtaining the double ratios one can try to parameterize these measurements in terms of the Collins functions. 
\subsection{Assumptions and input}
Assuming a Gaussian dependence on the intrinsic transverse momentum 
\begin{equation}
D_1(z,k_t)= \frac{D_1(z)}{\pi z^2}\exp(-k_t^2/\langle k_t^2\rangle)  \quad
\end{equation}
relative to the quark-antiquark axis it is possible to solve the convolutions of transverse momenta in the $A_0$, Eqs.~(\ref{eq:r0}),(\ref{eqn:sigma0}), asymmetries and to relate them to the $A_{12}$ asymmetries, Eqs.~(\ref{eq:r12def}),(\ref{eqn:sigma12}). As the Collins function has to obey the following positivity constraint \cite{alessandro}:
\begin{equation}
H_1^{\perp}(z,k_t) \frac{k_t}{zM} < D_1(z,k_t) \quad,
\end{equation}
the widths of the Gaussian distributions for the unpolarized fragmentation functions ($\langle k_t^2\rangle$) and the Collins fragmentation function ($\langle k_{tC}^2\rangle$) have to differ, since otherwise the constraint will not hold at sufficiently large transverse momenta.

Currently these Gaussian widths have to be taken as additional parameters. They are assumed to be universal between favored and disfavored fragmentation functions. In addition, $\mathrm{SU}(2)_f$ symmetry for u and d quarks is assumed for both types of fragmentation functions:
\begin{eqnarray}
H_1^{\perp,fav}(z,k_t)&:=& H_1^{\perp,u\rightarrow\pi^+}=H_1^{\perp,d\rightarrow\pi^-}\nonumber \\ &=&H_1^{\perp,\overline{d}\rightarrow\pi^+}=H_1^{\perp,\overline{u}\rightarrow\pi^-} \\
H_1^{\perp,dis}(z,k_t)&:=& H_1^{\perp,u\rightarrow\pi^-}=H_1^{\perp,d\rightarrow\pi^+}\nonumber \\ &=&H_1^{\perp,\overline{d}\rightarrow\pi^-}=H_1^{\perp,\overline{u}\rightarrow\pi^+}\,
\end{eqnarray}
and similarly for the unpolarized fragmentation functions $H_1^\perp \rightarrow D_1$. The strange quark fragmentation is ignored. Under this assumption one can rewrite the double ratio asymmetries entirely in terms of favored and disfavored unpolarized and Collins fragmentation functions. Applying the asymmetry definitions (see Eqn.~\ref{eqn:sigma12}) and integrating over the transverse momenta, the $A_{12}$ asymmetries thus become \cite{daniel3}($H_1^\perp(z_1)$ is abbreviated as $H_1$ and $D_1(z_1)$ as $D_1$ and equivalently for the $z_2$ dependent functions $H_2$ and $D_2$):
\begin{widetext}
\begin{eqnarray}
A^{UL}_{12} &=& \left\langle \frac{\sin^2\theta}{1+\cos^2\theta}\right\rangle \frac{\pi\langle k_{tC}^2 \rangle}{4M^2}%\nonumber \\
\Bigg[\frac{H_1^{fav}\overline{H}_2^{fav} + H_1^{dis}\overline{H}_2^{dis}}{D_1^{fav}\overline{D}_2^{fav} + D^{dis}_1\overline{D}_2^{dis}}- %\nonumber \\
\frac{H^{fav}_1\overline{H}_2^{dis} + H^{dis}_1\overline{H}_2^{fav}}{D^{fav}_1\overline{D}_2^{dis} + D_1^{dis}\overline{D}_2^{fav}}\Bigg] \\
A^{UC}_{12} &=& \left\langle \frac{\sin^2\theta}{1+\cos^2\theta}\right\rangle \frac{\pi\langle k_{tC}^2 \rangle}{4M^2}%\nonumber \\
\Bigg[\frac{H_1^{fav}\overline{H}_2^{fav} + H_1^{dis}\overline{H}_2^{dis}}{D_1^{fav}\overline{D}_2^{fav} + D^{dis}_1\overline{D}_2^{dis}}- %\nonumber \\
\frac{\left(H^{fav}_1 + H^{dis}_1\right)\left(\overline{H}_2^{fav}+\overline{H}_2^{dis}\right)}{\left(D^{fav}_1 + D_1^{dis}\right)\left(\overline{D}_2^{fav}+\overline{D}_2^{dis}\right)}\Bigg].
\end{eqnarray}
\end{widetext}

The $A_{0}$ asymmetries have a similar expression that differs by a factor $\pi/2$ \cite{daniel3}.

The Gaussian widths of the Collins functions have been fixed to be $\langle k_{tC}^2 \rangle /M^2 =2$, where $M$ is the pion mass. 
The unpolarized fragmentation functions in the denominator are taken either from \cite{kretzer}(Kretzer), \cite{hkns}(HKNS), \cite{marco}(DSS) or \cite{kkp}(KKP) at the scale of $Q^2=111$ GeV$^2$ in leading order. Since the latter do not contain explicit favored and disfavored fragmentation functions they were rescaled by $(1+z)/2$ for favored fragmentation and $(1-z)/2$ for disfavored fragmentation functions according to an assumption by \cite{feynman}. 
\subsection{Parameterization}
\begin{figure}[th]
\begin{center}
\includegraphics[width=8cm]{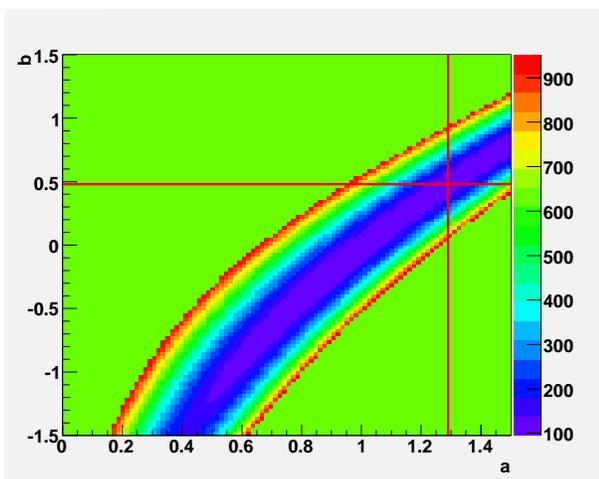}
\caption{\label{fig:chiparm}$\chi^2$ distribution as a function of the parameters $a$ and $b$ as specified in the text using the KKP parameterization of the unpolarized fragmentation functions for the $A_{0}$ data.  The $\chi^2$ values above $\chi^2_{max} : = 10 \times$ the minimum were set to 65\% of $\chi^2_{max}$ for better visibility. The minimum lies slightly below 100. The vertical and horizontal lines correspond to the parameters $a$ and $b$ of the lowest $\chi^2$.}
\end{center}
\end{figure}

Different parameterizations as a function of the fractional energy $z$ are possible for the Collins function. 
The simplest case takes Collins functions proportional to $z$ times the unpolarized fragmentation functions:
\begin{eqnarray}
H_1^{fav}(z) &=& a z D_1^{fav}(z) \\
H_1^{dis}(z) &=& b z D_1^{dis}(z) \quad.
\end{eqnarray}
In order to evaluate the sensitivity of such a parameterization we calculated the $\chi^2$ as a function of the two parameters $a$ and $b$. 
An example is shown in Fig.~\ref{fig:chiparm} for a  combination of $A_{0}^{UL}$ and $A^{UC}_{0}$ asymmetries using the KKP unpolarized fragmentation functions \cite{dave}. As the $\chi^2$ divided by the number of degrees of freedom (32-2) is around 3 a linear description in terms of two parameters seems to be too simple.  
The parameterization does not put stringent constraints on $a$ and $b$, because the minimum of $\chi^2$ occurs along a diagonal $a - b =$ const in the $a,b$ parameter plane. This can be explained by the quadratic nature of the measured asymmetries  in terms of the Collins fragmentation functions. Qualitatively speaking, a very large, favored Collins function can be compensated by an almost equally large disfavored Collins function. While it was hoped that the UC data could resolve this ambiguity it turns out that the sensitivity is limited and the favored over disfavored Collins function ratio $H_1^{fav}(z)/H_1^{dis}(z)$ remains inconclusive. However, if one restricts the ratio of the favored Collins function to the favored unpolarized fragmentation to be below unity then the opposite signs for the favored and disfavored Collins functions suggested by \cite{hermesprl} can be confirmed.

Several models predict the magnitude and the $z$ dependence of the Collins functions, although favored and disfavored fragmentation are in general not distinguished. Details can be found in the references \cite{alessandro0,alessandro2,alessandro3,leonard}.
While mostly the $[1/2]$ transverse momentum moments of the Collins functions are modeled these models are consistent with a bare Collins function rising with z relative to the unpolarized fragmentation function. They are thus also consistent with the rising asymmetries presented here.

In Ref.~\cite{peter} the previously published Belle data was used to parameterize the Collins functions. The results were then compared to the parameterizations obtained from the HERMES \cite{hermesprl} and COMPASS \cite{compassprl} data assuming a quark transversity distribution based on the chiral quark soliton model. Good agreement between the SIDIS and Belle data was found despite the different energy scales. In Ref.~\cite{alexei} a global fit based on the HERMES, COMPASS and the previously published Belle data has been performed to extract the first transversity distributions. However these parameterizations were still limited by the statistics of the previously published data set. With the newly obtained results it becomes possible to use the Collins functions to further constrain the quark transversity distribution.

\section{Summary and outlook}
We have presented a precise measurement of transverse spin asymmetries, which can be attributed to the product of a quark and an antiquark Collins function. The statistics has been improved by almost factor of 20 compared to the previously published results partially due to the inclusion of the data taken at the $\Upsilon (4S)$ resonance.  Different combinations of pion pairs exhibit significant, nonzero asymmetries. The systematic uncertainties in the measurements are understood and evaluated. The results of the measurements show significant spin-dependent effects in $e^+e^-\rightarrow q\bar{q}\ (q=u,d,s)$ fragmentation. In addition, assuming a Gaussian transverse momentum dependence of the Collins and unpolarized fragmentation functions, we studied a possible parameterization of the former. Due to the quadratic nature of the double ratios the difference between the favored and the disfavored Collins functions is still poorly determined.  
 However, suggestions based on semi-inclusive DIS data, that favored and disfavored Collins functions are both large and of opposite sign, are compatible with our measurement.
A global analysis, which was already applied using our previously published data set should further constrain both the Collins and the transversity functions.

\begin{table*}[bh]
\begin{ruledtabular}
\caption{\label{tab:dr}$A^{UL}_0$ and $A^{UL}_{12}$ values obtained from fits to pion double ratios as a function of {\it z}. The errors shown are statistical and systematic.}
\begin{tabular}{cccccccc}
$z_1 $&$\langle z_1 \rangle $&$z_2 $&$ \langle z_2 \rangle $ &$\left \langle \frac{\sin^2(\mathrm{ acos}\  \hat{n}_z)}{(1+\hat{n}_z^2) }\right\rangle $&$ \left\langle \frac{\sin^2 \theta_2 }{ (1+\cos^2 \theta_2)} \right\rangle $& $A^{UL}_0$&$A^{UL}_{12}$ \\ \hline \hline
$[0.2,0.3]$&$0.244$&$[0.2,0.3]$&$0.244$&$0.724$&$0.704$&$   0.0038\pm   0.0072\pm   0.0026$&$   0.0101\pm   0.0109\pm   0.0029$ \\
$[0.2,0.3]$&$0.244$&$[0.3,0.5]$&$0.377$&$0.727$&$0.708$&$   0.0204\pm   0.0031\pm   0.0027$&$   0.0300\pm   0.0048\pm   0.0031$ \\
$[0.2,0.3]$&$0.244$&$[0.5,0.7]$&$0.575$&$0.718$&$0.697$&$   0.0258\pm   0.0021\pm   0.0029$&$   0.0467\pm   0.0033\pm   0.0034$ \\
$[0.2,0.3]$&$0.244$&$[0.7,1.0]$&$0.779$&$0.719$&$0.701$&$   0.0414\pm   0.0013\pm   0.0033$&$   0.0609\pm   0.0020\pm   0.0037$ \\
$[0.3,0.5]$&$0.377$&$[0.2,0.3]$&$0.244$&$0.728$&$0.706$&$   0.0170\pm   0.0031\pm   0.0027$&$   0.0249\pm   0.0048\pm   0.0030$ \\
$[0.3,0.5]$&$0.377$&$[0.3,0.5]$&$0.377$&$0.730$&$0.710$&$   0.0265\pm   0.0024\pm   0.0029$&$   0.0462\pm   0.0036\pm   0.0034$ \\
$[0.3,0.5]$&$0.378$&$[0.5,0.7]$&$0.576$&$0.721$&$0.700$&$   0.0341\pm   0.0021\pm   0.0031$&$   0.0616\pm   0.0032\pm   0.0037$ \\
$[0.3,0.5]$&$0.379$&$[0.7,1.0]$&$0.778$&$0.722$&$0.704$&$   0.0630\pm   0.0016\pm   0.0041$&$   0.0770\pm   0.0025\pm   0.0042$ \\
$[0.5,0.7]$&$0.575$&$[0.2,0.3]$&$0.244$&$0.719$&$0.700$&$   0.0262\pm   0.0019\pm   0.0029$&$   0.0399\pm   0.0029\pm   0.0033$ \\
$[0.5,0.7]$&$0.576$&$[0.3,0.5]$&$0.378$&$0.721$&$0.705$&$   0.0349\pm   0.0018\pm   0.0031$&$   0.0567\pm   0.0027\pm   0.0036$ \\
$[0.5,0.7]$&$0.578$&$[0.5,0.7]$&$0.576$&$0.714$&$0.694$&$   0.0412\pm   0.0017\pm   0.0033$&$   0.1130\pm   0.0026\pm   0.0053$ \\
$[0.5,0.7]$&$0.578$&$[0.7,1.0]$&$0.780$&$0.715$&$0.697$&$   0.1069\pm   0.0017\pm   0.0059$&$   0.1191\pm   0.0031\pm   0.0055$ \\
$[0.7,1.0]$&$0.778$&$[0.2,0.3]$&$0.244$&$0.717$&$0.705$&$   0.0335\pm   0.0013\pm   0.0031$&$   0.0608\pm   0.0019\pm   0.0037$ \\
$[0.7,1.0]$&$0.779$&$[0.3,0.5]$&$0.379$&$0.718$&$0.703$&$   0.0524\pm   0.0025\pm   0.0037$&$   0.0796\pm   0.0039\pm   0.0042$ \\
$[0.7,1.0]$&$0.781$&$[0.5,0.7]$&$0.577$&$0.717$&$0.701$&$   0.0784\pm   0.0021\pm   0.0047$&$   0.1030\pm   0.0035\pm   0.0050$ \\
$[0.7,1.0]$&$0.783$&$[0.7,1.0]$&$0.780$&$0.715$&$0.705$&$   0.1525\pm   0.0150\pm   0.0086$&$   0.2063\pm   0.0225\pm   0.0091$ \\
\end{tabular}
\end{ruledtabular}
\end{table*}

\begin{table*}[ht]
\begin{ruledtabular}
\caption{\label{tab:dr2}$A^{UC}_0$ and $A^{UC}_{12}$ values obtained from fits to pion double ratios as a function of {\it z}. The errors shown are statistical and systematic.}
\begin{tabular}{cccccccc}
$z_1 $&$\langle z_1 \rangle $&$z_2 $&$ \langle z_2 \rangle $ &$ \left\langle \frac{\sin^2(\mathrm{acos}\  \hat{n}_z)}{(1+\hat{n}_z^2)} \right\rangle $&$ \left\langle \frac{\sin^2 \theta_2 }{ (1+\cos^2 \theta_2)} \right\rangle $& $A^{UC}_0$&$A^{UC}_{12}$ \\ \hline \hline
$[0.2,0.3]$&$0.244$&$[0.2,0.3]$&$0.244$&$0.724$&$0.704$&$  0.0016\pm   0.0030\pm   0.0012$&$   0.0029\pm   0.0030\pm   0.0026$ \\
$[0.2,0.3]$&$0.244$&$[0.3,0.5]$&$0.377$&$0.727$&$0.708$&$  0.0082\pm   0.0014\pm   0.0013$&$   0.0076\pm   0.0014\pm   0.0027$ \\
$[0.2,0.3]$&$0.244$&$[0.5,0.7]$&$0.575$&$0.718$&$0.697$&$  0.0103\pm   0.0009\pm   0.0013$&$   0.0118\pm   0.0009\pm   0.0029$ \\
$[0.2,0.3]$&$0.244$&$[0.7,1.0]$&$0.779$&$0.719$&$0.701$&$  0.0153\pm   0.0007\pm   0.0015$&$   0.0147\pm   0.0007\pm   0.0031$ \\
$[0.3,0.5]$&$0.377$&$[0.2,0.3]$&$0.244$&$0.728$&$0.706$&$  0.0068\pm   0.0014\pm   0.0013$&$   0.0061\pm   0.0014\pm   0.0027$ \\
$[0.3,0.5]$&$0.377$&$[0.3,0.5]$&$0.377$&$0.730$&$0.710$&$  0.0103\pm   0.0010\pm   0.0013$&$   0.0116\pm   0.0010\pm   0.0029$ \\
$[0.3,0.5]$&$0.378$&$[0.5,0.7]$&$0.576$&$0.721$&$0.700$&$  0.0126\pm   0.0009\pm   0.0014$&$   0.0148\pm   0.0009\pm   0.0031$ \\
$[0.3,0.5]$&$0.379$&$[0.7,1.0]$&$0.778$&$0.722$&$0.704$&$  0.0210\pm   0.0008\pm   0.0018$&$   0.0167\pm   0.0008\pm   0.0032$ \\
$[0.5,0.7]$&$0.575$&$[0.2,0.3]$&$0.244$&$0.719$&$0.700$&$  0.0103\pm   0.0008\pm   0.0013$&$   0.0099\pm   0.0008\pm   0.0028$ \\
$[0.5,0.7]$&$0.576$&$[0.3,0.5]$&$0.378$&$0.721$&$0.705$&$  0.0130\pm   0.0008\pm   0.0014$&$   0.0136\pm   0.0008\pm   0.0030$ \\
$[0.5,0.7]$&$0.578$&$[0.5,0.7]$&$0.576$&$0.714$&$0.694$&$  0.0137\pm   0.0007\pm   0.0014$&$   0.0252\pm   0.0007\pm   0.0038$ \\
$[0.5,0.7]$&$0.578$&$[0.7,1.0]$&$0.780$&$0.715$&$0.697$&$  0.0312\pm   0.0010\pm   0.0021$&$   0.0221\pm   0.0010\pm   0.0036$ \\
$[0.7,1.0]$&$0.778$&$[0.2,0.3]$&$0.244$&$0.717$&$0.705$&$  0.0121\pm   0.0007\pm   0.0015$&$   0.0145\pm   0.0007\pm   0.0031$ \\
$[0.7,1.0]$&$0.779$&$[0.3,0.5]$&$0.379$&$0.718$&$0.703$&$  0.0177\pm   0.0011\pm   0.0016$&$   0.0174\pm   0.0011\pm   0.0032$ \\
$[0.7,1.0]$&$0.781$&$[0.5,0.7]$&$0.577$&$0.717$&$0.701$&$  0.0226\pm   0.0011\pm   0.0018$&$   0.0191\pm   0.0011\pm   0.0034$ \\
$[0.7,1.0]$&$0.783$&$[0.7,1.0]$&$0.780$&$0.715$&$0.705$&$  0.0306\pm   0.0073\pm   0.0022$&$   0.0231\pm   0.0071\pm   0.0037$ \\ 
\end{tabular}
\end{ruledtabular}
\end{table*}
\begin{appendix}
\end{appendix}
\begin{acknowledgments}
The authors would like to thank D.~Boer for fruitful discussions on the theoretical aspects of the measurement. 

We thank the KEKB group for the excellent operation of the
accelerator, the KEK cryogenics group for the efficient
operation of the solenoid, and the KEK computer group and
the National Institute of Informatics for valuable computing
and SINET3 network support. We acknowledge support from
the Ministry of Education, Culture, Sports, Science, and
Technology of Japan and the Japan Society for the Promotion
of Science; the Australian Research Council and the
Australian Department of Education, Science and Training;
the National Natural Science Foundation of China under
contract No.~10575109 and 10775142; the Department of
Science and Technology of India; 
the BK21 program of the Ministry of Education of Korea, 
the CHEP SRC program and Basic Research program 
(grant No.~R01-2005-000-10089-0) of the Korea Science and
Engineering Foundation, and the Pure Basic Research Group 
program of the Korea Research Foundation; 
the Polish State Committee for Scientific Research; 
%-> remove for now: under contract No.~2P03B 01324; 
the Ministry of Education and Science of the Russian
Federation and the Russian Federal Agency for Atomic Energy;
the Slovenian Research Agency;  the Swiss
National Science Foundation; the National Science Council
and the Ministry of Education of Taiwan; and the U.S.\
Department of Energy and National Science Foundation.
\end{acknowledgments}

\newpage %Just because of unusual number of tables stacked at end
%\bibliography{LP_paper}% Produces the bibliography via BibTeX.

\end{document}